\documentclass[aps, prd, reprint, amsfonts,
  amssymb, amsmath, showpacs, preprintnumbers, letterpaper,
  nofootinbib]{revtex4-1}

\usepackage{braket,bm}
\usepackage[pass]{geometry} 
\usepackage{color}
\usepackage{mathrsfs,mathtools}
  \DeclareMathAlphabet{\mathpzc}{OT1}{pzc}{m}{it}
  
\newcommand{\D}{\mathrm{d}}

\begin{document}


\title{Green's function method for handling radiative effects on 
  false vacuum decay}

\author{Bj\"{o}rn Garbrecht}
\email{garbrecht@tum.de}

\author{Peter Millington}
\email{p.w.millington@tum.de}

\affiliation{Physik Department T70, James-Franck-Stra\ss e,\\
Technische Universit\"{a}t M\"{u}nchen, 85748 Garching, Germany}


\preprint{TUM-HEP-977-15}

\begin{abstract}

We introduce a Green's function  method for handling radiative effects
on  false   vacuum  decay.   In   addition  to  the   usual  thin-wall
approximation,  we  achieve  further simplification  by  treating  the
bubble  wall in  the planar  limit.  As  an application,  we take  the
$\lambda\Phi^4$ theory, extended with  $N$ additional heavier scalars,
wherein we  calculate analytically both the  functional determinant of
the quadratic  fluctuations about the classical  soliton configuration
and the first correction to the soliton configuration itself.

\end{abstract}

\maketitle


\section{Introduction}

The  association of  the still  recently-discovered $125\,  {\rm GeV}$
scalar particle \cite{Aad:2012tfa,  Chatrchyan:2012ufa} with the Higgs
boson of  the Standard Model (SM) places the stability of  the electroweak
vacuum under  question~\cite{Cabibbo:1979ay, Sher:1988mj, Sher:1993mf,
Isidori:2001bm}.  This instability, arising  at an energy scale around
$10^{11}\,   {\rm    GeV}$~\cite{EliasMiro:2011aa,   Degrassi:2012ry},
results   from  the   renormalization-group (RG) running   of  the   Higgs
self-coupling,  whose  value  is   driven  negative  by  contributions
dominated by top-quark  loops.  State-of-the-art calculations
suggest that the  electroweak vacuum is metastable,  having a lifetime
longer than the present  age of the Universe and lying  at the edge of
the     stable     region~\cite{EliasMiro:2011aa,     Degrassi:2012ry,
Alekhin:2012py,  Buttazzo:2013uya}, where  seemingly-small corrections
may have a material impact  upon predictions.  The uncertainty in such
predictions remains,  at present, dominated  by that of  the top-quark
pole  mass~\cite{Bezrukov:2012sa, Masina:2012tz}.  Even so, it has  been
suggested~\cite{Branchina:2013jra,                  Branchina:2014usa,
Branchina:2014rva, Lalak, Eichhorn:2015kea} that,  regardless of any improved  precision in the
experimental determination of the latter, the presence of Planck-scale
operators  may  weaken  the  claim  of  metastability.   Nevertheless,
having,  as yet,  no experimental  evidence of  additional stabilizing
physics between the electroweak and  Planck scales, it is provident to
consider approaches  to the  calculation of  tunneling rates  that can
consistently account for radiative corrections.

The  degree   of  vacuum
metastability provides  a strong criterion for  the phenomenological
viability of extensions to the SM. For example,  supersymmetric scenarios can be  ruled out if
the electroweak symmetry-breaking vacuum  decays into a color-breaking
one    in   a    timescale   shorter    than   the    age   of    the
Universe~\cite{Nilles:1982dy, AlvarezGaume:1983gj, Derendinger:1983bz,Kusenko:1996jn,Strumia:1996pr,
Chowdhury:2013dka,   Blinov:2013fta,    Camargo-Molina:2014pwa}.    In
addition,  transitions   between  vacua  can  also   occur  at  finite
temperature~\cite{Linde:1980tt,Linde:1981zj}.   In the  context  of early-Universe
cosmology, this  is of interest because  the corresponding first-order
phase    transitions   may    leave    behind   relic    gravitational
waves~\cite{Witten:1984rs, Kosowsky:1991ua, Caprini:2009fx}. Moreover,
such phase transitions  may turn out to be pivotal  for generating the
cosmic       matter-antimatter       asymmetry~\cite{Morrissey:2012db,
Chung:2012vg}.  As  a consequence of  these applications and  the wide
range of  phenomenological models, there  are now routine  methods for
computing transition   rates   at  both   vanishing   and   finite
temperature~\cite{Wainwright:2011kj, Camargo-Molina:2013qva}.

Vacuum  transitions  in  scalar  theories  can  be  described  in  the
following  way~\cite{Langer:1967ax,   Langer:1969bc,  Kobzarev:1974cp,
Coleman:1977py}. In the event that there are two non-degenerate vacua,
an  initially  homogeneous  system  lying in  the  false  vacuum  will
spontaneously  nucleate  bubbles  of   true  vacuum,  leading  to  the
production  of  domain  walls  or   ``kinks.''   The  latter  are  the
topological  solitons that  interpolate  between regions  of true  and
false  vacuum.  The  study  of these  ``solitary  wave'' solutions  to
non-linear equations of motion (see e.g.~Ref.~\cite{Scott:1973eg}) has
a long history~\cite{Dashen:1974cj, Polyakov:1974ek, Goldstone:1974gf,
Christ:1975wt, Jackiw:1977yn, Faddeev:1977rm}, and archetypal examples
of   such    field   configurations    arise   in    the   sine-Gordon
model~\cite{Coleman:1974bu,  Dashen:1975hd} and  the $\lambda  \Phi^4$
theory with tachyonic mass $m^2  < 0$.  The semi-classical~\cite{Coleman:1977py} and quantum~\cite{Callan:1977pt}
descriptions of false vacuum decay in the latter theory were presented
in  the  seminal  works by  Coleman  and  Callan (see also Ref.~\cite{Coleman:1978ae}). Early expansion on
these works included induced vacuum decay~\cite{Affleck:1979px} and the incorporation of gravity~\cite{Coleman:1980aw}.

In  order to decide  whether a  vacuum  configuration  is  unstable,
i.e.~whether  there exists  a  lowest-lying true  vacuum,  it is  often
necessary to account for the impact of radiative corrections. This    is   of
particular relevance when the appearance or disappearance of minima is entirely a radiative effect~\cite{Weinberg:1992ds}, such  as occurs for the Coleman-Weinberg (CW)
mechanism of spontaneous symmetry breaking~\cite{Coleman:1973jx} or in
symmetry  restoration  at  finite  temperature~\cite{Kirzhnits:1972ut,
Dolan:1973qd,  Weinberg:1974hy}. In phenomenological studies, this
is commonly done by calculating the tunneling rate from
the effective potential~\cite{Jackiw:1974cv,    Cornwall:1974vz} of  a \emph{homogeneous}
field   configuration~\cite{Camargo-Molina:2013qva,   Frampton:1976kf,
Frampton:1976pb}, which is subsequently promoted to a space-time-dependent configuration. This
practice is  problematic for two reasons. Firstly, the temporal and spatial inhomogeneity of the solitonic background is not fully taken into account~\cite{Surig:1997ne}. Secondly, in  the presence of  tachyonic instabilities,
e.g.~when there are non-convex regions in the tree-level
potential,
the perturbatively-calculated effective potential receives a  seemingly-pathological imaginary
part. The latter has been  shown~\cite{Weinberg:1987vp} to have a physical interpretation as a decay rate  for an
initially homogeneous field configuration
(see  also  Ref.~\cite{Einhorn:2007rv}). However, this subtlety can be circumvented by using constructions such as the coarse-grained effective action~\cite{Berges:1996ja,Berges:1996ib,Strumia:1998nf,Strumia:1998vd,Strumia:1998qq,Strumia:1999fq,Strumia:1999fv,Munster:2000kk} or, as is often employed in lattice simulations, the constraint effective potential~\cite{O'Raifeartaigh:1986hi} (cf.~Ref.~\cite{Alexandre:2012ht}). In addition, within the context of the SM, the standard RG improvement of the effective potential has recently been questioned~\cite{Gies:2014xha}. For the above reasons, it is reasonable to conclude that the use  of  the  effective
potential  to   calculate  transition   rates is neither
satisfactory nor justifiable  and it is desirable to consider alternative methods for determining quantum corrections, which can be applied to a wide range
of models that feature vacuum decay.

The  first  quantum corrections~\cite{Callan:1977pt}  to  the tunneling rate  are  those
arising from the functional  determinant over the quadratic fluctuations
about   the  classical   soliton  configuration.    In  the   case  of
one-dimensional operators, these determinants  may be calculated using
the  Gel'fand-Yaglom   theorem~\cite{Gelfand:1959nq},  which   may  be
generalized  to higher  dimensions in  the case  of radially-symmetric
operators~\cite{Baacke:1993ne,Dunne:2006ct,Dunne:2007rt}.
General numerical techniques may then be obtained~\cite{Baacke:2003uw,
Dunne:2005rt}  for calculating  tunneling rates  beyond the  so-called
thin-wall approximation, in which the width of the bubble wall is much
smaller than  its radius. These  approaches have also been  applied to
radially-separable      Yang-Mills     backgrounds~\cite{Dunne:2006ac,
Dunne:2007mt}         and          scenarios         in         curved
spacetime~\cite{Dunne:2006bt}. Alternatively,  as  we  will  employ,  the
functional determinant may be calculated by means of the so-called heat kernel
method (see  e.g.~Refs.~\cite{Diakonov:1983dt,Diakonov:1984xj,Konoplich:1987yd, Vassilevich:2003xt}),  based  upon the Schwinger proper-time representation and zeta function regularization~\cite{Hawking:1976ja}. Previously, this approach has been used to derive approximate analytic results for the one-loop fluctuation determinant beyond the thin-wall approximation~\cite{Munster:1999hr}, as well as one-loop corrections to sphaleron rates~\cite{Carson:1989rf,Carson:1990jm,Carson:1990wp}. The latter have also been calculated by direct integration of the Green's function~\cite{Baacke:1993jr,Baacke:1993aj, Baacke:1994ix, Baacke:1994bk,Baacke:2008zx}.

Recently, it  has been shown  that properties of  topological solitons
may  be  studied  non-perturbatively  using Monte  Carlo  and  lattice
simulations       by      considering       correlation      functions
directly~\cite{Rajantie:2009bk,                       Rajantie:2010tb,
Rajantie:2010fw}. Other authors have  proposed methods for calculating
quantum    corrections   based    upon   functional    renormalization
techniques~\cite{Alexandre:2007ci}.

In this article, we derive an \emph{analytic} result for the Green's function
of the $\lambda \Phi^4$ theory in the background of the classical kink
solution.   Within  the  thin-   and  planar-wall  approximations,  we
illustrate that  this Green's  function may be  used to  determine \emph{analytically} the
leading quantum  corrections to both the  semi-classical bounce action
and the  kink solution  itself, thereby allowing us to compute the tunneling rate at the two-loop level, while isolating its diagrammatic interpretation. The  latter calculation  is performed
within the  context of  a toy  model extended  with an  additional $N$
heavier  scalars, where the parametric dependence on $N$ allows the identification of a concrete example in which the calculated two-loop corrections dominate over the neglected higher-loop corrections. We illustrate  that  the problem  of
calculating  these radiative  corrections  may be  reduced  to one  of
solving   one-dimensional   ordinary    differential   equations   and
integrals. Thus,  we anticipate  that this methodical  development may
have  numerical  applications in  the  study  of  the decay  rates  of
radiatively-generated metastable vacua, such  as occur in the massless
CW model~\cite{Coleman:1973jx} or the Higgs potential of
the SM. Similar methods based upon Green's function techniques have been used previously to determine self-consistent bounce solutions \emph{numerically} in the Hartree approximation of the pure $\lambda\Phi^4$ theory in both two and four dimensions~\cite{Bergner:2003au,Bergner:2003id,Baacke:2004xk,Baacke:2006kv}.	

The  remainder   of  this  article   is  organized  as   follows.   In
Sec.~\ref{sec:classb},  we review  the  calculation, \textit{\`{a}  la}
  Coleman  and  Callan~\cite{Coleman:1977py, Callan:1977pt},  of  the
classical  ``bounce''  configuration,  describing  the  semi-classical
tunneling  rate  between two  quasi-degenerate  vacua  and its  first
quantum corrections.   In Sec.~\ref{sec:greens}, we outline  a Green's
function method for the evaluation  of the functional determinant over
the quantum fluctuations about the classical bounce, making comparison
with existing calculations.  Subsequently, in Sec.~\ref{sec:corrb}, we
illustrate that this Green's function  method may be used to calculate
analytically and  self-consistently the  first quantum  corrections to
the bounce itself. In Sec.~\ref{sec:conc}, we conclude our discussions
and highlight potential applications and future directions. Finally, a
number  of   mathematical  appendices  are  included,   outlining  the
technical    details    of     the    calculations    summarized    in
Secs.~\ref{sec:greens} and~\ref{sec:corrb}.


\section{Semi-classical bounce}
\label{sec:classb}

We  consider  a   real  scalar  field  $\Phi   \equiv  \Phi(x)$,  with
four-dimensional Euclidean  Lagrangian $\mathcal{L}  = (\partial_{\mu}
\Phi)^2/2 + U$ and classical potential
\begin{equation}
  \label{eq:pot}
  U \ = \ \frac{1}{2!} \, m_{\Phi}^2 \Phi^2  + \frac{g}{3!} \, \Phi^3 
  + \frac{\lambda}{4!} \, \Phi^4 + U_0 \; .
\end{equation}
The mass  squared is $m_{\Phi}^2  = -  \, \mu^2 <  0$, $g$ is  of mass
dimension one,  $\lambda$ is  dimensionless, $U_0$  is a  constant and
$\partial_{\mu}  \equiv  \partial  /  \partial  x_{\mu}$  denotes  the
derivative with respect to the Euclidean spacetime coordinate $x_{\mu}
\equiv  (\mathbf{x},x_4)$. Throughout,  we  omit  spacetime and  field
arguments for notational convenience when no ambiguity results.

\begin{figure}
  \centering
  \includegraphics[scale=1]{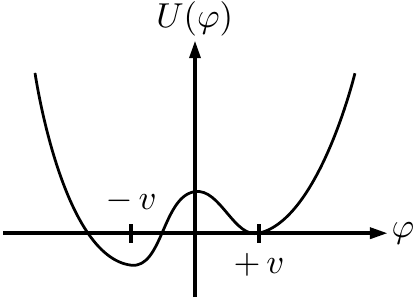} \hspace{0.15em}
  \includegraphics[scale=1]{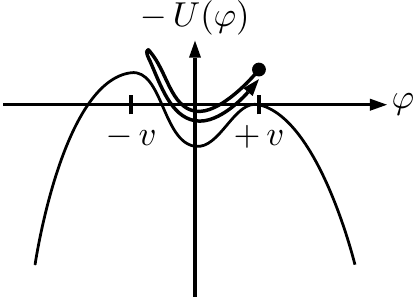}
  \caption{The classical  potential $U$ (left panel)  and the inverted
potential $-\,U$ (right panel). The  arrow (right panel) indicates the
trajectory     of     the     ``bounce''     in     imaginary     time
$\tau$.\label{fig:potential}}
\end{figure}

The classical  potential in Eq.~\eqref{eq:pot} has  non-degenerate minima
at
\begin{equation}
  \varphi \ = \ v_{\pm} \ = \ \pm \, v \, \bigg(1 
  + \frac{\bar{v}^2}{v^2}\bigg)^{\!\frac{1}{2}} - \bar{v} \; ,
\end{equation}
as depicted  in Fig.~\ref{fig:potential}  (left panel), where  we have
defined $v  = \sqrt{6  \mu^2 /  \lambda}$ and  $\bar{v} =  (3 g)  / (2
\lambda)$. The separation of  the minima $\Delta v = v_+ -  v_- = 2 d$
and the  difference in their  energy densities  $\Delta U =  U_{v_+} -
U_{v-} = 2 \varepsilon$ may be written in terms of the parameters
\begin{subequations}
\begin{align}
  d &= v \, \bigg(\! 1 + \frac{\bar{v}^2}{v^2}\bigg)^{\!\!\frac{1}{2}} \!
  \approx v\;, \\ \frac{\varepsilon}{d} &= \frac{gv^2}{6} \bigg(\! 1
  + 3 \, \frac{\bar{v}^2}{v^2}\bigg) \approx \frac{gv^2}{6} \; ,
\end{align}
\end{subequations}
where  the approximations  are valid  in  the limit  $v \gg  \bar{v}$,
i.e.~$g^2 / \mu^2 \ll 8 \lambda /  3$. For $g \to 0$, $\varepsilon \to
0$ and  the minima at  $\varphi = \pm \,  v$ become degenerate,  as we
would expect.

Finally, the constant  $U_0$ is chosen so that  the potential vanishes
in the  false vacuum  at $\varphi  = +  \, v$,  requiring $U_0  = (\mu
v/2)^{2} - g  v^3 / 6$ and thus  giving the barrier height to  be $h =
U_0 + 2 \varepsilon \approx (\mu v / 2)^2 + \varepsilon$ and $U(-\,v)=-\,gv^3/3$.

The  semi-classical  probability  for   tunneling  between  the  false
($\varphi = + \, v$) and true ($\varphi = - \, v$) vacua and its first
quantum corrections were described in the seminal works by Coleman and
Callan \cite{Coleman:1977py, Callan:1977pt}, which  we now review. The
classical equation of motion
\begin{equation}
  \label{eq:classeom}
  - \, \partial^2 \varphi + U'(\varphi) \ = \ 0\;,
\end{equation}
where $'$ denotes the derivative  with respect to the field $\varphi$,
is analogous  to that of  a particle moving in  a potential $-  \, U$.
The boundary conditions  of the ``bounce'' are  $\varphi|_{x_4 \to \pm
\infty}  = +  \, v$  and $\dot{\varphi}|_{x_4  = 0}  \ =  \ 0$,  where
$\dot{}$  denotes   the  derivative  with  respect   to  $x_4$.  These
correspond to  a particle initially  at $+  \, v$ rolling  through the
valley in $-  \, U$, reaching a turning point  close to $-\,v$, before
rolling back to $+ \,  v$, see Fig.~\ref{fig:potential} (right panel).
Finally, in order  to ensure that the action of  the bounce is finite,
we require $\varphi|_{|\mathbf{x}| \to \infty} = + \, v$.

Translating    to    four-dimensional   hyperspherical    coordinates,
Eq.~\eqref{eq:classeom} takes the form
\begin{equation}
  \label{eq:classeomrad}
  - \frac{\D^2}{\D r^2} \, \varphi
  - \frac{3}{r} \, \frac{\D}{\D r} \, \varphi
  + U'(\varphi) \ = \ 0 \; ,
\end{equation}
where $r^2  = \mathbf{x}^2  + x_4^2$.  The boundary  conditions become
$\varphi|_{r \to \infty} =  + \, v$ and $\D \varphi / \D  r|_{r = 0} =
0$, where the latter ensures that  the solution is well defined at the
origin. Thus, the  bounce corresponds to a  four-dimensional bubble of
some radius $R$, which separates the false vacuum ($\varphi = + \, v$)
outside from the true vacuum inside ($\varphi = - \, v$). Analytically
continuing to Minkowski spacetime ($x_4 = i x_0$), the $O(4)$ symmetry
of the bounce becomes an $SO(1,3)$ symmetry, with the bubble expanding
along the hyperbolic trajectory $R^2 \ = \ \mathbf{x}^2 - c^2 t^2$.

The bounce action is
\begin{equation}
  \label{eq:B}
  B \ = \ \int \! \D^4x \; \bigg[ \frac{1}{2}
  \bigg( \frac{\mathrm{d\varphi}}{\mathrm{d}r} \bigg)^{\!2}
  + U(\varphi) \bigg] \; ,
\end{equation}
which can be written as $B = B_{\rm surface} + B_{\rm vacuum}$, where
\begin{subequations}
  \begin{align}
    \label{eq:Bsurface}
    B_{\rm surface} \ & = \ 2 \pi^2 R^3 \int_{- \, v}^{+ \, v} \!
    \D \varphi \; \frac{\mathrm{d\varphi}}{\mathrm{d}r} \; ,
    \\
    B_{\rm vacuum} \ & = \ 2 \pi^2 \int_{0}^{R} \! \D r \;
    r^3 U(- \, v)
  \end{align}
\end{subequations}
are the contributions  from the surface tension of the  bubble and the
energy    of   the    true    vacuum,    respectively.   In    writing
Eq.~\eqref{eq:Bsurface},   we   have   used    the   fact   that   for
$\partial_{\mu} \varphi \neq 0$, i.e.~for $r \sim R$, we may show that
the bounce $\varphi$ satisfies the virial theorem
\begin{equation}
  \bigg(\frac{\D \varphi}{\D r}\bigg)^{\! 2} - 2 U(\varphi) \ = \ 0 \; ,
\end{equation}
i.e.~it is the configuration of zero total energy density. Notice that
there is  no contribution to  the bounce action [Eq.~\eqref{eq:B}] from
the  exterior  of  the  bubble,  since the  choice  of  the  potential
Eq.~\eqref{eq:pot}, viz.~$U_0$, ensures that the false vacuum has zero
energy density.

In the thin-wall approximation, we may safely neglect the damping term
in  Eq.~\eqref{eq:classeomrad} and  the  contribution  from the  cubic
self-interaction $g\varphi^3$, as  will be the case  for the remainder
of   this    article.   We    then   obtain   the    well-known   kink
solution~\cite{Dashen:1974cj}
\begin{equation}
  \varphi(r) \ = \ v \tanh \big[\gamma ( r - R )\big] \; ,
\end{equation}
with $\gamma = \mu  / \sqrt{2}$. The radius $R$ of  the bubble is then
obtained by extremizing the bounce action Eq.~\eqref{eq:B}, that is by
minimizing the  energy difference between  the surface tension  of the
bubble and the true vacuum. This gives
\begin{equation}
  R \ = \ \frac{12 \gamma}{g v} \; .
\end{equation}

By considering  the invariance  of the bounce  action in Eq.~\eqref{eq:B}
under general coordinate transformations,  i.e.~$\varphi \to \varphi +
x_{\mu} \partial_{\mu} \varphi$, we may show that
\begin{equation}
  \label{eq:fullbounceact}
  B \ = \ \frac{1}{2} \pi^2 R^3 \int_{- \, v}^{+ \, v} \! \D \varphi \, 
  \frac{\D \varphi}{\D r} \; .
\end{equation}
This is to say that
\begin{equation}
  B_{\rm vacuum} \ = \ - \, \frac{3}{4} \, B_{\rm surface} \; , 
\end{equation}
in which case we find
\begin{equation}
  B = \ - \, \frac{1}{3} B_{\rm vacuum} \
  = \ - \, \frac{\pi^2}{6} \, R^4 \, U(- \, v) \
  = \ \frac{8 \pi^2 R^3 \gamma^3}{\lambda} \; .
\end{equation}

The decay rate of the false vacuum, i.e.~the probability per unit time
for  the nucleation  of  a  bubble of  true  vacuum,  has the  generic
form~\cite{Coleman:1977py, Callan:1977pt}
\begin{equation}
  \label{eq:tunnelprob}
  \varGamma \ = \ A V e^{-\,B / \hbar} \; .
\end{equation}
Here,  $V$ is  the three-volume  within  which the  bounce may  occur,
arising  from integrating  over  the  center of  the  bounce, and  $A$
contains the  quantum corrections to  the classical bounce  action $B$
that are the subject of the remainder of this article.

The tunneling probability in Eq.~\eqref{eq:tunnelprob} may be obtained
from the path integral
\begin{equation}
  \label{eq:path}
  Z[0] \ = \ \int \! [ \D \Phi ] \; e^{- \, S[\Phi] / \hbar} \; ,
\end{equation}
via
\begin{equation}
  \label{eq:tunnelrel}
  \varGamma \ = \ 2 | \mathrm{Im} \, Z[0] | / T \; ,
\end{equation}
where $T$ is the Euclidean time of the bounce.

In order  to evaluate  the functional integral  over $\Phi$,  we first
expand  around  the  classical  bounce $\varphi$,  whose  equation  of
motion~\eqref{eq:classeom} is obtained from
\begin{equation}
  \label{eq:deltaS}
  \frac{\delta S[\Phi]}{\delta \Phi} \bigg|_{\Phi \, = \, \varphi} \
  = \ 0 \; .
\end{equation}
Writing  $\Phi =  \varphi  + \hbar^{1/2}  \hat{\Phi}$, where  the
factor  of  $\hbar^{1/2}$  is   written  explicitly  for  bookkeeping
purposes, we find
\begin{equation}
  S[\Phi] \ = \ S[\varphi] + \, \frac{\hbar}{2} \! \int \! \D^4 x \;
  \hat{\Phi}(x) \, G^{-1}(\varphi;x) \, \hat{\Phi}(x)
  + \mathcal{O}(\hbar^{3/2}) \; ,
\end{equation}
where $S[\varphi] \equiv B$ is the classical bounce action and
\begin{equation}
  \label{eq:Gminusdef}
  G^{-1}(\varphi;x) \ \equiv \
  \frac{\delta^2 S[\Phi]}{\delta \Phi^2(x)}\bigg|_{\Phi \, = \, \varphi}
  \ = \ - \, \Delta^{\! (4)} + U''(\varphi; x) \; ,
\end{equation}
in which $\Delta^{\! (4)}$ is the four-dimensional Laplacian.

Before  proceeding  to perform  the  functional  integration over  the
quadratic fluctuations about the bounce, we must consider the spectrum
of   the  operator   $G^{-1}(\varphi;x)$,   which   is  not   positive
definite.    By    differentiating     the    equation    of    motion
Eq.~\eqref{eq:classeom} with  respect to $x_{\mu}$ and  comparing with
the eigenvalue equation
\begin{equation}
  \big( - \Delta^{(4)} + U''(\varphi) \big) \phi_{\{n\}} \
  = \ \lambda_{\{n\}} \phi_{\{n\}} \; ,
\end{equation}
it is  straightforward to show  that there exist four  zero eigenmodes
$\phi_{\mu} =  \mathcal{N} \partial_{\mu} \varphi$, transforming  as a
vector of $SO(4)$  and resulting from the  translational invariance of
the   bounce.    The    normalization   $\mathcal{N}$   follows   from
Eq.~\eqref{eq:fullbounceact}, since
\begin{equation}
  \int \! \D^4 x \; \phi_{\mu}^* \phi_{\nu} \ = \ \frac{1}{4} \,
  \mathcal{N}^2 \delta_{\mu \nu}  \! \int \! \D^4 x \; 
  \big( \partial_{\lambda} \varphi \big)^2 \
  = \ \mathcal{N}^2 B \, \delta_{\mu \nu} \; .
\end{equation}
Thus, $\phi_{\mu} = B^{-1/2} \partial_{\mu} \varphi$.

Differentiating  Eq.~\eqref{eq:classeomrad} with  respect  to $r$  and
subsequently  setting $r  = R$  in  those terms  originating from  the
damping term, we can show that  there also exists a discrete eigenmode
$\phi_0 = B^{-1/2} \partial_r \varphi$. This eigenmode transforms as a
scalar of $SO(4)$, corresponding to dilatations of the classical bounce solution, and has the negative eigenvalue
\begin{equation}
  \lambda_0 \ = \ \frac{1}{B} \frac{\delta^2\!B}{\delta R^2} \
  = \ - \, \frac{3}{R^2} \; .
\end{equation}
It is  this lowest mode that  is responsible for the  path integral in
Eq.~\eqref{eq:path}   obtaining  the   non-zero   imaginary  part   in
Eq.~\eqref{eq:tunnelrel}~\cite{Coleman:1987rm}.

Alternatively,  we  may  solve  the  eigenvalue  problem  directly  in
hyperspherical coordinates (see  Appendix~\ref{app:greens}), by making
the substitution $\phi_{\{n\}} = \phi_{nj}/r^3$.\footnote{We note that
this      substitution     differs      from     that      used     in
Ref.~\cite{Callan:1977pt}.}  Neglecting  the damping term  and setting
$r = R$ in the centrifugal potential, we obtain the eigenspectrum
\begin{equation}
  \label{eq:eigenspectrum}
  \lambda_{nj} \ = \ \gamma^2 \big( 4 - n^2 \big)
  + \frac{j(j + 2) - 3}{R^2} \; .
\end{equation}
The radial  parts of  the eigenfunctions  are the  associated Legendre
polynomials    of    the    first     kind    and    of    order    2,
i.e.~$P_{2}^{n}(\varphi/v)$.   Thus,  demanding  normalizability,  the
quantum number $n$ is restricted to the set $\{1,2\}$.

From  Eq.~\eqref{eq:eigenspectrum},  we  see that  the  negative  mode
corresponds to $\lambda_0  = \lambda_{20}$ ($n = 2$, $j  = 0$) and the
zero modes  correspond to $\lambda_{21}$  ($n =  2$, $j =  1$), having
degeneracy  $(j  +  1)^2  =  4$.   The  lowest  two  positive-definite
eigenvalues are $\lambda_{10} =  2 \gamma^2 - 3 / R^2$ ($n  = 1$, $j =
0$) and $\lambda_{11} = 2 \gamma^2$ ($n = 1$, $j = 1$).  Thus, for $R$
large,  the   ``continuum''  of  positive-definite  modes   begins  at
$\lambda_{10}     \approx     \lambda_{11}      =     2     \gamma^2$,
cf.~Ref.~\cite{Konoplich:1987yd}.

In  order   to  perform   the  functional   integral  over   the  five
negative-semi-definite  discrete modes,  we expand  $\hat{\Phi}  =
\sum_{i  =  0}^{4}a_i  \phi_i  + \phi_+$,  where  $\phi_+$  comprises  the
continuum of positive-definite eigenmodes. The functional measure then
becomes
\begin{equation}
  [ \D \Phi ] \ = \ [ \D \phi_+ ] \prod_{i = 0}^4
  (2 \pi \hbar)^{- 1/2} \D a_i \; .
\end{equation}

The functional integral over the four  zero eigenmodes ($i = 1, \dots,
4$) is traded  for an integral over the collective  coordinates of the
bounce~\cite{Gervais:1974dc}  (see   Appendix~\ref{app:zeromode})  and
yields a factor
\begin{equation}
  V T \bigg( \frac{B}{2 \pi \hbar} \bigg)^{\!2} \; .
\end{equation}
The integral  over the negative eigenmode  ($i = 0$) may  be performed
using the method of steepest descent, giving an overall factor of $- i
| \lambda_0   |^{-1/2}    /   2$.   Here,   the    overall   sign   is
unphysical~\cite{Callan:1977pt} and depends on  the choice of analytic
continuation,    thereby    justifying    the    modulus    sign    in
Eq.~\eqref{eq:tunnelrel}.

Finally,  the  Gaussian  integral   over  the  continuum  of  positive
eigenmodes $\phi_+$ may be performed in the usual manner and we obtain
\begin{equation}
  \label{eq:pathinted}
  i Z[0] \ = \ e^{- B / \hbar} \left|
  \frac{\lambda_0 \, \mathrm{det}^{(5)} \, G^{-1}(\varphi)}
       {\frac{1}{4} (VT)^2 \big(\frac{B}{2 \pi \hbar}\big)^{\!4}
       (4 \gamma^2)^5 \, \mathrm{det}^{(5)} \, G^{-1}(v)}
  \right|^{-\tfrac{1}{2}} \; ,
\end{equation}
in   which   $\mathrm{det}^{(5)}$,   cf.~Ref.~\cite{Konoplich:1987yd},
denotes  the  determinant  calculated   only  over  the  continuum  of
positive-definite  eigenmodes,  i.e.~omitting  the zero  and  negative
eigenmodes, whose contributions are  included explicitly. In addition,
we  have   normalized  the  determinant   to  that  of   the  operator
$G^{-1}(v)$,   evaluated   in    the   false   vacuum.    Substituting
Eq.~\eqref{eq:pathinted}  into Eq.~\eqref{eq:tunnelrel},  we find  the
tunneling rate per unit volume
\begin{align}
  \label{eq:tun}
  & \varGamma / V \ = \ \bigg( \frac{B}{2 \pi \hbar}\bigg)^{\!2}
  (2 \gamma)^5 | \lambda_0 |^{-\tfrac{1}{2}} \exp \bigg[ 
    - \frac{1}{\hbar} \bigg( B + \hbar B^{(1)} \bigg) \bigg] \; ,
\end{align}
where
\begin{equation}
  \label{eq:B1}
  B^{(1)} \ = \ \frac{1}{2} \, \mathrm{tr}^{(5)}
  \Big( \ln G^{-1}(\varphi) - \ln G^{-1}(v) \Big) \; ,
\end{equation}
contains  the one-loop  corrections  from  the quadratic  fluctuations
around the classical bounce.  Here, $\mathrm{tr}^{(5)}$ indicates that
we are to trace over only the positive-definite eigenmodes.


\section{Green's function method}
\label{sec:greens}

In this section, we outline the  derivation of the Green's function of
the operator  in Eq.~\eqref{eq:Gminusdef}.  The technical  details are
included for completeness  in Appendix~\ref{app:greens}. Subsequently,
we use this Green's function to evaluate the functional determinant in
Eq.~\eqref{eq:pathinted}  and  obtain  the correction  from  quadratic
fluctuations.  In  addition, we calculate analytically the  tadpole contribution to
the effective equation  of motion and point out that this  may be used
to calculate the first quantum corrections to the bounce.

We have the inhomogeneous Klein-Gordon equation
\begin{equation}
  \label{eq:phiKG}
  \big( \! - \Delta^{\! (4)} + U''(\varphi; x) \big) G(\varphi; x, x') \
  = \ \delta^{(4)}(x - x') \; ,
\end{equation}
where  $\delta^{(4)}(x  - x')$  is  the  four-dimensional Dirac  delta
function. Working  in hyperspherical coordinates and  writing $x_{\mu}
\!     \!      \!     {}^{(}{}'{}^{)}     =      r     {}^{(}{}'{}^{)}
\mathbf{e}_{r{}^{(}{}'{}^{)}}$,  where $\mathbf{e}_{r{}^{(}{}'{}^{)}}$
are  four-dimensional  unit  vectors,  the  Green's  function  may  be
expanded as
\begin{equation}
  \label{eq:greenexp}
  G(\varphi; x, x') \ = \ \frac{1}{2 \pi^2} \sum_{j = 0}^{\infty}(j + 1)
  G_{j}(\varphi; r, r') U_j(\cos \theta) \; ,
\end{equation}
where $\cos \theta =  \mathbf{e}_r \cdot \mathbf{e}_{r'}$ and $U_j(z)$
are   the   Chebyshev   polynomials    of   the   second   kind   (see
Appendix~\ref{app:greens}). The radial  functions $G_j(r, r')$ satisfy
the inhomogeneous equation
\begin{align}
  \label{eq:radialdif}
  &\bigg[ - \frac{\D^2}{\D r^2} - \frac{3}{r} \frac{\D}{\D r}
  + \frac{j(j + 2)}{r^2}
  \nonumber \\
  & \qquad \qquad
  + U''(r) \bigg] G_j(\varphi; r, r')\
  = \ \frac{\delta(r - r')}{r'^3} \; .
\end{align}

For the thin wall, we safely  neglect the damping term and approximate
the centrifugal term by $j(j + 2) / R^2$. For self-consistency of this
approximation,  we  also  replace  the discontinuity  on  the  rhs  of
Eq.~\eqref{eq:radialdif} with $\delta(r - r')/R^3$.  For  generality of
notation in what follows, it is then convenient to define
\begin{align}
  G(u, u', m) \ & \equiv \ R^3 G_j(\varphi; r, r') \; ,
\end{align}
being a function only of the normalized bounce
\begin{equation}
\label{phi:u}
  u{}^{(}{}'{}^{)} \ \equiv \ \frac{\varphi(r{}^{(}{}'{}^{)})}{v} \
  = \ \tanh \big[\gamma (r{}^{(}{}'{}^{)} - R)\big]
\end{equation}
and the parameter
\begin{equation}
  m \ = \ 2 \bigg( 1 + \frac{j(j + 2)}
    {4 \gamma^2 R^2} \bigg)^{\! \tfrac{1}{2}} \; .
\end{equation}
The full Green's function may then be written
\begin{align}
  \label{eq:greennotedef}
  G(\varphi; x, x') \ & \equiv \ G(u, u', \theta)
  \nonumber \\
  & = \ \frac{1}{2 \pi^2 R^3} \sum_{j = 0}^{\infty} (j + 1)
  U_j(\cos \theta)G(u, u', m) \; .
\end{align}

With the above approximations,  the lhs of Eq.~\eqref{eq:radialdif} is
of   P\"{o}schl-Teller   form~\cite{Poschl:1933zz},   having   general
solutions that  may be expressed  in terms of the  associated Legendre
functions (see  Appendix~\ref{app:greens}). We  are then able  to find
the full analytic solution
\begin{align}
  \label{eq:Guupm}
  & G(u, u', m) \ = \ \frac{1}{2 \gamma m}\bigg[ \vartheta(u - u') 
  \bigg(\frac{1 - u}{1 + u}\bigg)^{\! \! \tfrac{m}{2}} \! 
  \bigg(\frac{1 + u'}{1 - u'}\bigg)^{\! \! \tfrac{m}{2}}
  \nonumber \\ 
  & \quad \times \bigg( \! 1 - 3 \,
  \frac{(1 - u)(1 + m + u)}{(1 + m)(2 + m)} \bigg)
  \nonumber \\
  & \quad \times \bigg( \! 1 - 3 \,
  \frac{(1 - u')(1 - m + u')}{(1 - m)(2 - m)} \bigg)
  + (u \leftrightarrow u') \bigg] \; ,
\end{align}
where $\vartheta(z)$ is the generalized unit-step function.

Taking  the coincidence  limit  $u  = u'$,  $\theta  =  0$, the  local
contribution  to the  Green's function  $G(u)  \equiv G(u,  u, 0)$  in
Eq.~\eqref{eq:greennotedef} takes the form
\begin{equation}
  \label{eq:Gu}
  G(u) \ = \ \frac{1}{2 \pi^2 R^3}
  \sum_{j = 0}^{\infty} (j + 1)^2 G(u, m) \; ,
\end{equation}
where
\begin{align}
  \label{eq:Gum}
  & G(u, m) \ \equiv \ G(u, u, m)
  \nonumber \\
  & \ = \ \frac{1}{2 \gamma m}\bigg[ 1 + 3
  \big( 1 - u^2 \big) \sum_{n = 1}^2
  \frac{(-1)^n (n - 1 - u^2) }{m^2 - n^2} \bigg] \; .
\end{align}
In Eq.~\eqref{eq:Gum},  the summation over  $n = 1, 2$  corresponds to
the contributions from the  two towers of positive-definite eigenmodes
of       the       operator      $G^{-1}(\varphi;       x)$, see
Eq.~\eqref{eq:eigenspectrum}.

For  $R$ large,  we  may  approximate the  summation  over  $j$ by  an
integral  over a  continuous variable  $k \sim  \frac{j +  1}{R}$ (see
Appendix~\ref{app:greens}). In which case, we obtain
\begin{equation}
  \label{eq:Grk2}
  G(u) \ = \ \frac{1}{2 \pi^2} \!
  \int_{0}^{\infty} \D k \; k^2 \, G(u, m) \; ,
\end{equation}
with
\begin{equation}
  \label{eq:continuum}
  m \ = \ 2 \, \bigg( 1 +
  \frac{k^2}{4 \gamma^2} \bigg)^{\! \tfrac{1}{2}} \; .
\end{equation}

\begin{figure}
\centering
  \includegraphics[scale=1]{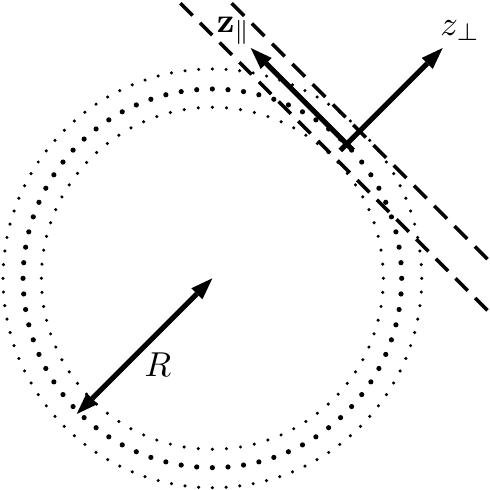}
  \caption{Schematic representation of  the planar-wall approximation,
    illustrating the  alignment of the  three-dimensional hypersurface
    and associated coordinate system.\label{fig:planar}}
\end{figure}

The  continuum limit  described above  is entirely  equivalent to  the
so-called planar-wall approximation. Therein,  for $R$ large, we align
a set  of coordinates  $(z_{\perp}, \mathbf{z}_{\parallel})$  with the
bubble wall,  as shown in  Fig.~\ref{fig:planar}. We may  then Fourier-transform  with respect  to  the coordinates  $\mathbf{z}_{\parallel}$
that   lie   within   the  three-dimensional   wall,   introducing   a
three-momentum $\mathbf{k}$, i.e.
\begin{equation}
  G(\varphi; x, x') \ = \ \! \int \! \!
  \frac{\D^3 \mathbf{k}}{(2 \pi)^3} \; 
  e^{i\mathbf{k}\cdot(\mathbf{z}_{\parallel} - \mathbf{z}'_{\parallel})} \, 
  G(\varphi; z, z', \mathbf{k}) \; ,
\end{equation}
where  we have  let $z  = z_{\perp}$  for notational  convenience. The
three-momentum-dependent   Green's   function   $G(\varphi;   z,   z',
\mathbf{k})$ satisfies the inhomogeneous Klein-Gordon equation
\begin{equation}
  \label{eq:planwallKG}
  \big( - \partial_z^2 + k^2 + U''(\varphi; z) \big)
  G(\varphi; z, z', \mathbf{k}) \ = \ \delta(z - z') \; .
\end{equation}
We may then show straightforwardly that
\begin{equation}
  G(\varphi; z, z' , \mathbf{k}) \ = \ G(u, u', m) \; ,
\end{equation}
where $G(u,  u', m)$  is as defined  in Eq.~\eqref{eq:Guupm}  with $m$
given by  Eq.~\eqref{eq:continuum}. This planar-wall  approximation is
employed for the remainder of this article.


\subsection{Quantum-corrected bounce}

Before making use of the Green's function calculated in the preceding
section,   we  first   derive   the  equation   of   motion  for   the
quantum-corrected  bounce.  This  calculation was  first suggested  by
Goldstone  and Jackiw~\cite{Goldstone:1974gf}  and,  in the  following
sections, we  will illustrate that,  within the thin-  and planar-wall
approximations, it may be completed analytically.

The one-particle irreducible (1PI) effective action~\cite{Jackiw:1974cv} is
given by the Legendre transform
\begin{equation}
  \label{effact}
  \Gamma[\phi] \ = \ - \, \hbar \, \ln Z[J] + 
  \int \! \D^4 x \; J(x) \phi(x) \; ,
\end{equation}
where
\begin{equation}
\phi(x)\ =\ \hbar\,\frac{\delta \ln Z[J]}{\delta J(x)}
\end{equation}
is a functional of the source $J(x)=\frac{\delta \Gamma[\phi]}{\delta \phi(x)}$ and
\begin{equation}
  \label{ZofJ}
  Z[J] \, = \int \! [ \D \Phi ] \; \exp \! 
  \bigg[ - \frac{1}{\hbar} \bigg( 
  S[\Phi] - \! \int \! \D^4 x \; J(x) \Phi(x) \bigg) \bigg] \; .
\end{equation}

In order to obtain the quantum corrections to the bounce $\varphi$, we
wish  to evaluate  the  functional integral in Eq.~\eqref{ZofJ} by  expanding around  the
configuration $\varphi^{(1)}$,  which is  the solution to  the quantum
equation of motion
\begin{equation}
  \label{eq:gamdif}
  \frac{\delta \Gamma[\phi]}{\delta \phi} \bigg|_{\phi = \varphi^{(1)}}
  \ = \ 0 \; .
\end{equation}
Here,  the   superscript  ``$(1)$''  indicates   that  $\varphi^{(1)}$
contains the first  quantum corrections to $\varphi$.  It follows from
Eq.~\eqref{eq:gamdif}  that   $\varphi^{(1)}$  cannot   extremize  the
classical action in the absence of the source $J$, i.e.
\begin{equation}
  \label{eq:actdif}
  \frac{\delta S[\Phi]}{\delta \Phi(x)}\bigg|_{\Phi \, = \, \varphi^{(1)}} \
  = \ J(x) \ \neq \ 0 \; .
\end{equation}

Writing  $\Phi =  \varphi^{(1)} +  \hbar^{1/2} \hat{\Phi}^{(1)}$,
where  the factor  of $\hbar^{1/2}$  is again  written explicitly  for
bookkeeping, we proceed as in Sec.~\ref{sec:classb}, expanding
\begin{align}
  & S[\Phi] \ = \ S[\varphi^{(1)}] + \hbar^{1/2} \! \int \! \D^4 x \; 
  J(x) \, \hat{\Phi}^{(1)}(x)
  \nonumber \\
  & \quad + \frac{\hbar}{2} \int \! \D^4 x \; \hat{\Phi}^{(1)}(x) \,
  G^{-1}(\varphi^{(1)}; x) \, \hat{\Phi}^{(1)}(x) + \cdots \; ,
\end{align}
where
\begin{equation}
  \label{eq:G1minusdef}
  G^{-1}(\varphi^{(1)}; x) \ \equiv \
  \frac{\delta^2 S[\Phi]}{\delta \Phi^2(x)}\bigg|_{\Phi \, = \, \varphi^{(1)}} \
  = \ - \, \Delta^{(4)} + U''(\varphi^{(1)}; x) \; .
\end{equation}
We now write $\varphi^{(1)} = \varphi + \hbar\delta\varphi$ and expand about the quadratic fluctuations evaluated around the classical bounce $\varphi$. Thus,
in  performing the  functional  integral, we  consider the  same
spectrum of  negative and zero eigenmodes  as in Sec.~\ref{sec:classb}. Finally, by expanding the effective action $\Gamma[\phi]$ in Eq.~\eqref{effact} around $\varphi^{(1)}=\phi-\hbar\delta\varphi$ (see Ref.~\cite{Carrington:2004sn}), we obtain
\begin{align}
  \label{eq:effact}
  & \Gamma[\varphi^{(1)}] \ = \ S[\varphi^{(1)}]
  + \frac{i \pi \hbar}{2}+\hbar^2B^{(2)}{}'[\varphi]
  \nonumber \\
  & \qquad + \frac{\hbar}{2} \ln \left|
  \frac{\lambda_0 \, \mathrm{det}^{(5)} \, G^{-1}(\varphi)}
    {\frac{1}{4} (VT)^2 \big( \frac{B}{2 \pi \hbar} \big)^4
    (4 \gamma^2)^5 \mathrm{det}^{(5)} \, G^{-1}(v)} \right| + \cdots \;,
\end{align}
where
\begin{align}
B^{(2)}{}'[\varphi]\ &= \ \frac{1}{2}\int\!\D^4 x\;\delta\varphi(x)\nonumber\\&\quad\times \frac{\delta}{\delta \varphi^{(1)}(x)}\,\ln\frac{\mathrm{det}^{(5)}G^{-1}(\varphi^{(1)})}{\mathrm{det}^{(5)}G^{-1}(v)}\Bigg|_{\varphi^{(1)}=\varphi}\;.
\end{align}

Functionally  differentiating Eq.~\eqref{eq:effact}  with respect  to $\varphi^{(1)}$, we obtain the equation of motion for the corrected bounce
\begin{equation}
  \label{eq:eom1}
  - \, \partial^2 \varphi^{(1)}(x) + U_{\mathrm{eff}}'(\varphi^{(1)};x) \
  = \ 0 \; ,
\end{equation}
where
\begin{equation}
  \label{eq:Ueffprime}
  U_{\mathrm{eff}}'(\varphi^{(1)}; x) \ \equiv \ U'(\varphi^{(1)}; x)
  + \hbar\Pi(\varphi;x) \varphi(x) \; ,
\end{equation}
containing the tadpole contribution
\begin{equation}
  \label{eq:tadpole}
  \Pi(\varphi;x)\ =\ \frac{\lambda}{2} \, G(\varphi; x, x)\;.
\end{equation}
Comparing  the  functional  derivative of  Eq.~\eqref{eq:effact}  with
Eqs.~\eqref{eq:gamdif}   and~\eqref{eq:actdif},  we   see  that   this
evaluation of the  effective action is self-consistent so  long as the
source
\begin{equation}
  \label{eq:Jdef}
  J(x) \ = \ - \,\hbar\Pi(\varphi;x) \varphi(x) \; ,
\end{equation}
which is, as expected, non-vanishing.

We may show that the correction  to the classical bounce $\delta \varphi$
satisfies the equation of motion
\begin{equation}
  \label{eq:deltaphieom}
  G^{-1}(\varphi; x) \, \delta \varphi(x) \
  = \ - \, \Pi(\varphi;x) \varphi(x) \; .
\end{equation}

The corrected bounce  action $S[\varphi^{(1)}]$ contains contributions
at order  $\hbar^2$. Specifically,
\begin{align}
  \label{eq:secondexp}
  & S[\varphi^{(1)}] \ = \ S[\varphi]
  \nonumber \\
  & \quad + \, \frac{\hbar^2}{2} \! \int \! \D^4 x \; \delta \varphi(x)
  \, G^{-1}(\varphi; x) \, \delta \varphi(x)
  + \mathcal{O}(\hbar^{3}) \; ,
\end{align}
where        we       have        used       Eqs.~\eqref{eq:deltaS} and \eqref{eq:Gminusdef}.     Thus,     using     Eq.~\eqref{eq:deltaphieom},  we may write
\begin{equation}
  S[\varphi^{(1)}] \ = \ B + \hbar^2 B^{(2)} \; ,
\end{equation}
where
\begin{equation}
  \label{eq:B2}
  B^{(2)} \ = \ - \, \frac{1}{2} \! \int \! \D^4 x \; 
  \varphi(x) \Pi(\varphi;x) \delta \varphi(x) \; .
\end{equation}

Hence, we obtain the tunneling rate per unit volume
\begin{align}
  \varGamma / V \ &= \  2 \, | \mathrm{Im} \, e^{-\Gamma[\varphi^{(1)}]/\hbar} | / (VT)
   \nonumber\\ &=\ \bigg( \frac{B}{2 \pi\hbar} \bigg)^{\! 2} (2 \gamma)^5
  | \lambda_0 |^{- \frac{1}{2}}\nonumber \\ &\quad \times \exp \! \bigg[ - \frac{1}{\hbar}
  \bigg( B + \hbar B^{(1)} + \hbar^2 B^{(2)}+\hbar^2B^{(2)}{}' \bigg) \bigg] \, ,
\end{align}
where  $B$  is  the  classical  bounce  action;  $B^{(1)}$,  given  in
Eq.~\eqref{eq:B1}, contains the corrections from quadratic fluctations
about the classical bounce; and $B^{(2)}$, given in Eq.~\eqref{eq:B2},
contains the contribution arising from  the quantum corrections to the
bounce itself.  We note that
\begin{equation}
B^{(2)}{}'=-2B^{(2)}\;,
\end{equation}
such that the $\mathcal{O}(\hbar)$ corrections to the quadratic fluctuations flip the sign of the contribution to the bounce action from the $\mathcal{O}(\hbar)$ corrections to the bounce itself.


\subsection{Tadpole contribution}

We will now  proceed to calculate explicitly  the tadpole contribution
appearing in Eq.~\eqref{eq:tadpole}.

Introducing an ultraviolet cutoff $\Lambda$,  the $k$ integral can be
performed in Eq.~\eqref{eq:Grk2}, and we obtain
\begin{align}
  G(u) \ & = \ \frac{\gamma^2}{8 \pi^2} \bigg[
  \frac{\Lambda^2}{\gamma^2} + 2 - \big( 1 - 3 u^2 \big)
  \ln \frac{\gamma^2}{\Lambda^2}
  \nonumber \\ 
  & \qquad - \sqrt{3} \pi u^2 \big( 1 - u^2 \big) \bigg] \; .
\end{align} We choose to define the  physical mass and coupling in the
homogeneous non-solitonic background.\footnote{It is natural to define
the renormalized quantities  in the false vacuum, since  this is where
the physical  measurements of  these quantities  are performed.  If it
were the  case that such  measurements were  taking place in  the true
vacuum, or indeed within the wall itself, then the decay rate would be
of  little  concern.}   The  renormalization conditions  are  then  as
follows:
\begin{subequations}
  \label{eq:renormconds}
  \begin{gather}
    \frac{\partial^2 U_{\rm eff}(\varphi)}{\partial \varphi^2}
    \bigg|_{\varphi = v} \
    = \ - \, \mu^2 + \frac{\lambda}{2} \, v^2 \
    = \ 2 \mu^2 \; ,
    \\
    \frac{\partial^4 U_{\rm eff}(\varphi)}{\partial \varphi^4}
    \bigg|_{\varphi = v} \ = \ \lambda \; ,
  \end{gather}
\end{subequations}
where   $U_{\mathrm{eff}}$    is   the    CW  effective
potential~\cite{Coleman:1973jx}.  The  resulting   mass  and  coupling
counterterms are
\begin{align}
  \delta m^2 & = - \, \frac{\lambda \gamma^2}{16 \pi^2}
  \bigg(\frac{\Lambda^2}{\gamma^2} - 
  \ln \frac{\gamma^2}{\Lambda^2} - 31 \bigg) \; ,
  \\
  \delta \lambda & = - \, \frac{3 \lambda^2}{32 \pi^2}
  \bigg( \ln \frac{\gamma^2}{\Lambda^2} + 5 \bigg) \; .
\end{align}
We then arrive at the renormalized tadpole correction
\begin{align}
  \label{eq:phitad}
  & \Pi^R(u) \ = \ \frac{\lambda}{2} \, G(u)
  + \delta m^2 + \frac{2\gamma^2}{\lambda} \delta \lambda \, u^2 
  \nonumber \\
  & \qquad = \ \frac{3 \lambda \gamma^2}{16 \pi^2}
  \bigg[6 + \big(1 - u^2\big)
  \bigg(5 - \frac{\pi}{\sqrt{3}} \, u^2 \bigg) \bigg] \; .
\end{align}


\subsection{Functional determinant}

We   may  calculate   the  traces   appearing  in   the  exponent   of
Eq.~\eqref{eq:tun}, which arise from the functional determinant of the
operator $G^{-1}(\varphi)$  in Eq.~\eqref{eq:pathinted}, by  using the
heat kernel method (see e.g.~Ref.~\cite{Vassilevich:2003xt}). Specifically,
the trace may be written in the form
\begin{equation}
  \mathrm{tr}^{(5)} \ln G^{-1}(\varphi;x) \ = \ - \int\! \D^4 x \! 
  \int_{0}^{\infty} \! \frac{\D \tau}{\tau} \;
  K(\varphi; x, x | \, \tau) \; .
\end{equation}
The heat kernel $K(\varphi;  x, x' | \, \tau)$ is  the solution to the
heat-flow equation
\begin{equation}
  \label{hflow}
  \partial_{\tau} K(\varphi; x, x' | \, \tau) \
  = \ G^{-1}(\varphi; x) K(\varphi; x, x' | \, \tau)
\end{equation}
and satisfies the condition $K(\varphi; x, x' | \, 0) = \delta^{(4)}(x
- x') \, $.

It is convenient to work in terms of the Laplace transform of the heat
kernel
\begin{equation}
  \mathcal{K}(\varphi; x, x' | s) \ = \ \int_0^{\infty}\!\D \tau \; 
  e^{s\tau} \, K(\varphi; x, x' | \, \tau) \; ,
\end{equation}
which is the solution to
\begin{equation}
  \big( - \partial^2 + s + U''(\varphi; x) \big)
  \mathcal{K}(\varphi; x, x' | \, s) \ = \ \delta^{(4)}(x - x') \; .
\end{equation}
In the planar-wall approximation, we take
\begin{equation}
  \label{eq:LapK}
  \mathcal{K}(\varphi; x, x' | \, s) \ = \ \! 
  \int \! \! \frac{\D^3\mathbf{k}}{(2\pi)^3}\;e^{i\mathbf{k}\cdot(\mathbf{z}_{\parallel}-\mathbf{z}_{\parallel}')}\,
  \mathcal{K}(\varphi; z, z', \mathbf{k} | \, s) \; ,
\end{equation}
where $\mathcal{K}(\varphi; z, z', \mathbf{k} | \, s)$ satisfies
\begin{equation}
  \label{eq:LapKplanwall}
  \big( - \partial_z^2 + k^2 + s + U''(\varphi; z) \big)
  \mathcal{K}(\varphi; z, z', \mathbf{k} | \, s) \
  = \ \delta(z - z') \; .
\end{equation}

Comparing Eq.~\eqref{eq:LapKplanwall}  with Eq.~\eqref{eq:planwallKG},
we  see that  $\mathcal{K}(\varphi;  z,  z', \mathbf{k}  |  \, s)$  is
nothing   other  than   the  Green's   function  $G(u,   u',  m)$   in
Eq.~\eqref{eq:Guupm} with  the replacement $k^2  \to k^2 + s$  in $m$,
see Eq.~\eqref{eq:continuum}. Thus, we may write
\begin{align}
  \label{eq:B1explicit}
  B^{(1)} \ & = \ - \, \frac{1}{2} \! \int_{0}^{\Lambda} \! 
  \D k \; k^2 \! \int_{0}^{\infty} \! \frac{\D \tau}{\tau}
  \int_{0}^{\infty} \! \! \D r \; r^3 \, 
  \mathcal{L}^{-1}_s[\widetilde{G}(u, m)](\tau) \; ,
\end{align}
where we have defined
\begin{equation}
  \label{eq:Gtilde}
  \widetilde{G}(u, m) \ = \ G(u, m) - G(1, m)
\end{equation}
and
\begin{equation}
  \mathcal{L}^{-1}_s[\widetilde{G}(u, m)](\tau) \
  = \ \int_{\mathcal{C}} \! \frac{\D s}{2\pi i} \;
  e^{-s\tau} \, \widetilde{G}(u, m)
\end{equation}
is  the   inverse  Laplace  transform   with  respect  to   $s$,  with
$\mathcal{C}$ indicating the Bromwich contour.

We   may   perform    the   integrals   in   Eq.~\eqref{eq:B1explicit}
analytically, proceeding  in order  from right  to left  and beginning
with the inverse Laplace transform.  We then obtain the unrenormalized
correction to the bounce action
\begin{align}
  B^{(1)} \ = \ - \, B \, \bigg( \frac{3 \lambda}{16 \pi^2} \bigg)
  \bigg( \frac{\pi}{3 \sqrt{3}} +
  \frac{\Lambda^2}{\gamma^2}
  + \ln \frac{\gamma^2}{\Lambda^2} \bigg) \; .
\end{align}
The technical  details of  the relevant  integrations are  included in
Appendix~\ref{app:greens}. Adding the counterterm
\begin{align}
  \label{eq:deltaB}
  \delta B^{(1)} \ & = \ \int \! \D^4 x \; 
  \bigg( \frac{1}{2!} \delta m^2 \big( \varphi^2 - v^2 \big)
  + \frac{1}{4!} \delta \lambda \big( \varphi^4 - v^4 \big) \bigg)
  \nonumber \\
  & = \ B \, \bigg( \frac{3 \lambda}{16 \pi^2} \bigg)
  \bigg( \frac{\Lambda^2}{\gamma^2}
  + \ln \frac{\gamma^2}{\Lambda^2} - 21 \bigg) \; ,
\end{align}
we obtain the final renormalized result
\begin{align}
  \label{eq:B1final}
  B^{(1)} \ = \ - \, B \, \bigg( \frac{3 \lambda}{16 \pi^2} \bigg)
  \bigg( \frac{\pi}{3 \sqrt{3}} + 21 \bigg) \; .
\end{align}
In Appendix~\ref{app:greens},  we reproduce this result  by the method
presented  in  Ref.~\cite{Konoplich:1987yd}.\footnote{Using  the  same
renormalization   conditions    as   in   Eq.~\eqref{eq:renormconds},
Ref.~\cite{Konoplich:1987yd} finds (in the notation employed here)
  \begin{equation*}
    B^{(1)} \ = \ - \, B \, \bigg( \frac{3 \lambda}{16 \pi^2} \bigg)
   \bigg( \frac{\pi}{3 \sqrt{3}} + \frac{50}{3} \bigg) \; .
  \end{equation*}
  Repeating   the  analysis  presented
  therein,  as  outlined  in Appendix~\ref{app:greens},  we  find  a
  result  in agreement  with  Eq.~\eqref{eq:B1final} reported  here,
  suggesting a numerical error in the factor of 50 above.}


\section{Radiative corrections to the bounce}
\label{sec:corrb}

We now  discuss an example of  the role played by  loop corrections to
the  bounce  itself. Within  the  perturbation  expansion, one  should
expect that  these lead to  second-order corrections to  the classical
action of  the soliton simply  because the  latter is evaluated  for a
stationary path.  There are, however, important situations, in which all
one-loop  contributions  must be  resummed  in  order to  capture  the
leading quantum corrections to the action. Examples include situations
where the symmetry-breaking minima of the potential emerge radiatively
through the CW  mechanism~\cite{Coleman:1973jx}.  In the
absence  of  a soliton,  this  implies  that the  classical  solution,
i.e.~the homogeneous expectation  value of the field, has  to be found
consistently by minimizing the one-loop  effective potential as a {\it
function} of the field expectation value itself. Analogously, in order
to  find the  decay  rate of  the  false vacuum,  the  bounce must  be
computed consistently from  the one-loop effective action,  which is a
{\it functional} of  the bounce itself. The methods  presented in this
article  reduce  the problem  of  tunneling  in  radiatively-generated
potentials  to  one-dimensional  ordinary differential  equations  and
integrals.  It is  anticipated that  it should  be possible  to derive
numerical solutions in future work.
  
For  the purpose  of illustration,  however, we  remain herein  on the
ground  of  analytic  and  perturbative approximations.  In  order  to
enhance  the  corrections to  the  bounce  compared to  other  quantum
effects that appear at second  order in perturbation theory, we extend
the  model in  Eq.~\eqref{eq:pot}  with $N$  copies  of an  additional
scalar field $\chi$ by adding to the Lagrangian the terms
\begin{align}
  {\cal L}\chi \ = \ \sum \limits_{i = 1}^N \left\{
  \frac{1}{2!} (\partial_{\mu} \chi_i)^2
  + \frac{1}{2!} m_\chi^2 \chi_i^2
  + \frac{\lambda}{4} \Phi^2 \chi_i^2  \right\} \, .
\end{align}
Here, we  have chosen the  coupling $\lambda$  to be identical  to the
self-coupling  of $\Phi$  for the  sake of  simplicity in  the Green's
function  of the  $\chi$  fields.  Since  $\braket{\chi_i}  = 0$,  the
additional  scalars  do  not  impact  upon  the  classical  bounce  in
Sec.~\ref{sec:classb}  or the  discussion of  the Green's  function in
Sec.~\ref{sec:greens}.

The Klein-Gordon equation for $\chi_i$ takes the form
\begin{equation}
  \bigg[ - \partial^2 + m_{\chi}^2 + 
  \frac{\lambda}{2} \varphi^2 \bigg] S(\varphi; x, x') \
  = \ \delta^{(4)}(x - x') \; .
\end{equation}
Comparing with that of $\Phi$ in Eq.~\eqref{eq:phiKG}, we see that the
Green's function $S(u, u', m)$  may be obtained straightforwardly from
$G(u, u', m)$ in Eq.~\eqref{eq:Guupm} by making the replacement
\begin{equation}
  m \to \sqrt{6} \, \bigg( 1 + \frac{k^2 + m_{\chi}^2}
      {6 \gamma^2} \bigg)^{\! \tfrac{1}{2}} \; .
\end{equation}

The  renormalized  tadpole  contribution  from  each  $\chi_i$  field,
integrated over the three-momentum $\mathbf{k}$, is given by
\begin{equation}
  \label{eq:SR}
  \Sigma^{R}(u) \ = \ \frac{\lambda \gamma^2}{8 \pi^2} \,
  \frac{\gamma^2}{m_{\chi}^2} \, 
  \big[ 72 + \big( 1 - u^2 \big) \big( 40 - 3 u^2 \big) \big] \; ,
\end{equation}
where we have assumed $m_\chi^2 \gg \gamma^2$ for simplicity. The full
form  of  $S(u)$  and  the   relevant  counterterms  are  provided  in
Appendix~\ref{app:Scalc}.

The renormalized  correction to the classical  bounce $\delta \varphi$
is governed by the equation of motion
\begin{align}
  \bigg[  \frac{\D^2}{\D r^2} + \mu^2 -
  \frac{\lambda}{2} \varphi^2 \bigg]
  \delta \varphi \ = \ 
  \Big( \Pi^R(u) + N \Sigma^R(u) \Big) \, \varphi \; ,
\end{align}
cf.~Eq.~\eqref{eq:deltaphieom}. We  obtain the solution by  making use
of the Green's  function $G(u, u', 2)  \equiv G(u, u', m)|_{k  \, = \,
  0}$, writing
\begin{align}
  \label{sol:corr:Green}
  \delta \varphi(u) & = - \, \frac{v}{\gamma} \! 
  \int_{-1}^1 \! \D u' \, \frac{u' G(u, u', 2)}{1 - u'^2} \,
  \Big( \Pi^R(u') + N \Sigma^R(u') \Big) \, ,
\end{align}
where we have used Eq.~\eqref{phi:u} in order to substitute $\varphi$.

\begin{figure}
  \includegraphics[scale=0.6]{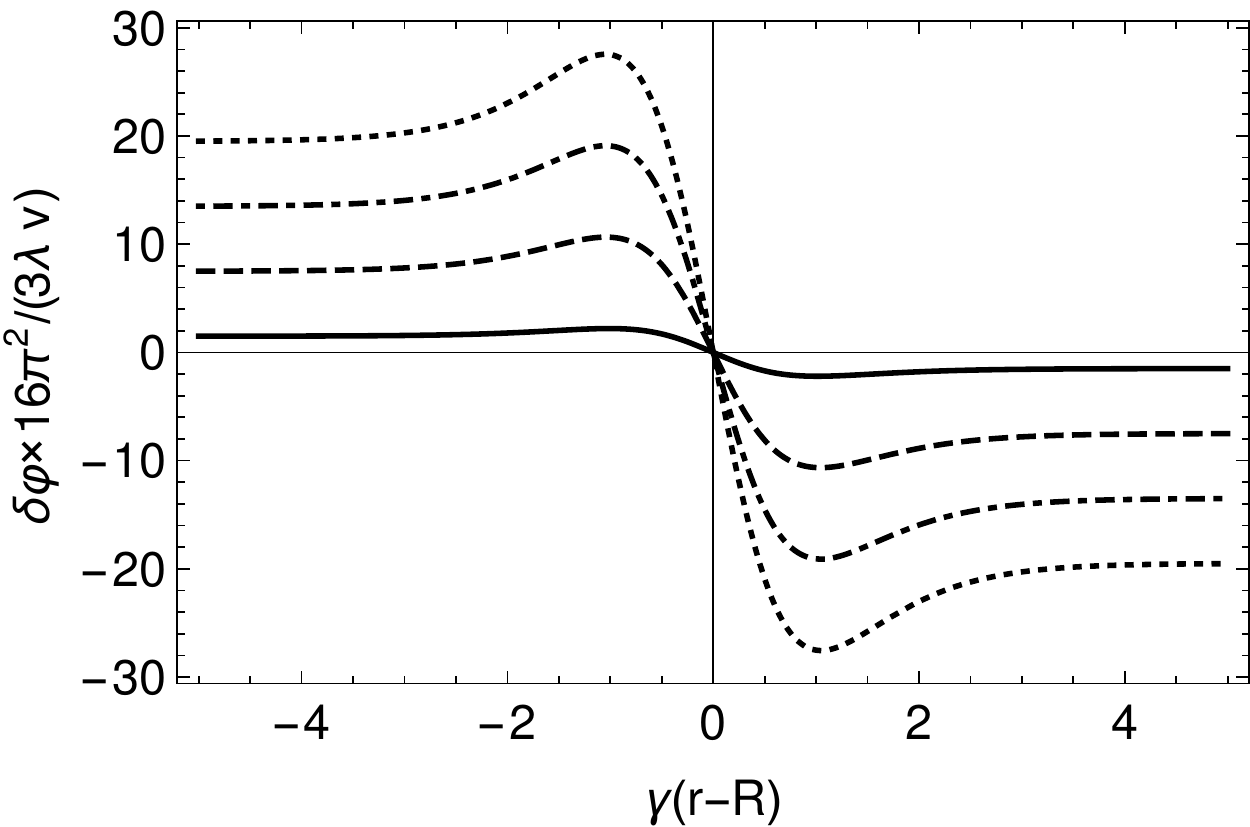}
  \caption{The correction to the bounce $\delta \varphi$ as a function
of   $\gamma(r-R)$   for   $N\gamma^2/m_{\chi}^2=0$   (solid),   $0.5$
(dashed), $1$ (dash-dotted) and $1.5$ (dotted).\label{fig:deltaphi}}
\end{figure}

\begin{figure}
  \includegraphics[scale=0.6]{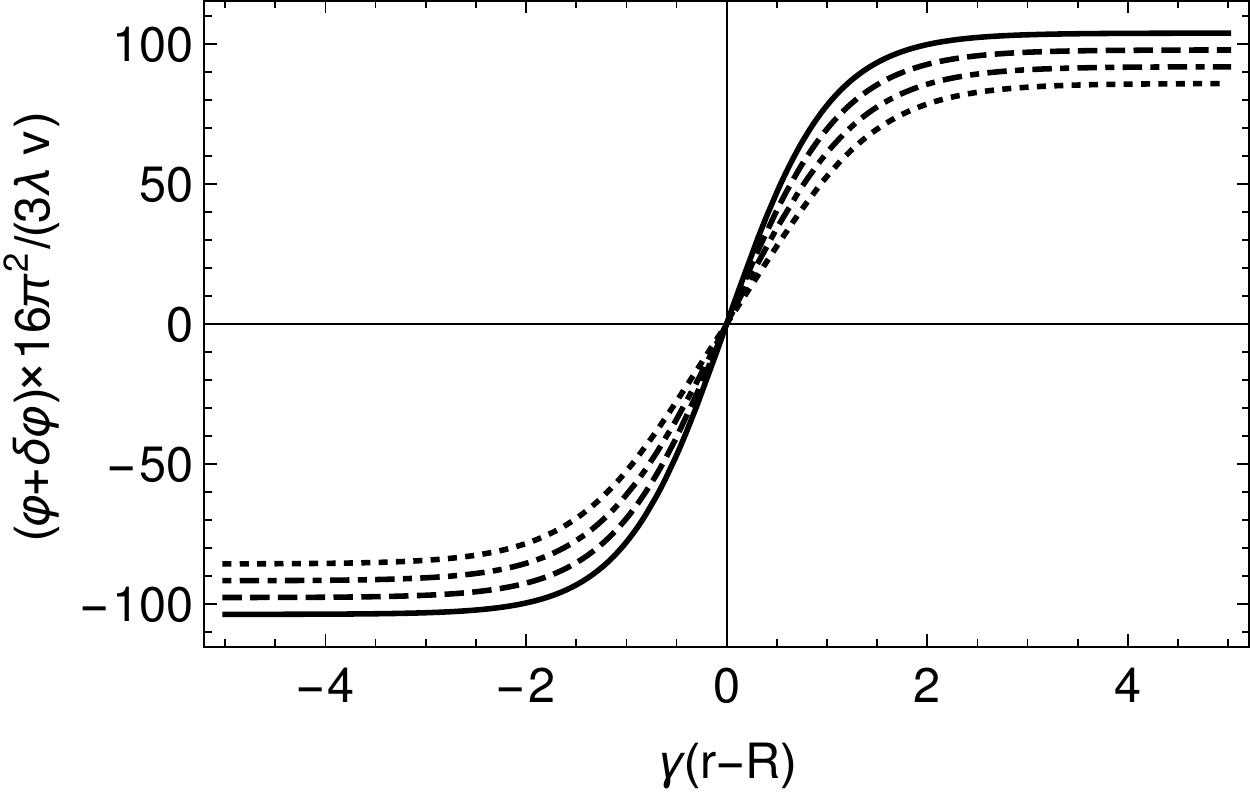}\\
  \vspace{1em}
  \includegraphics[scale=0.6]{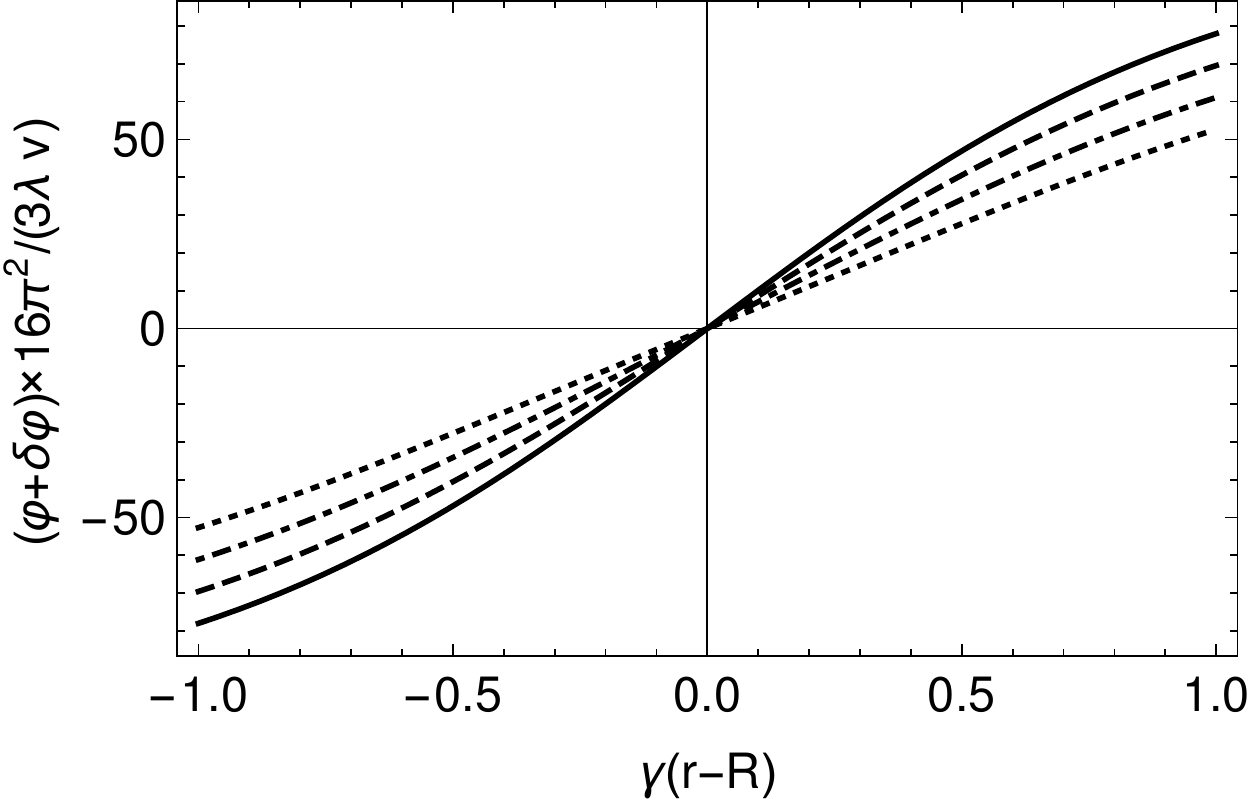}
  \caption{The  corrected  bounce  $\varphi  + \delta  \varphi$  as  a
function of $\gamma(r-R)$  for $N\gamma^2/m_{\chi}^2=0$ (solid), $0.5$
(dashed), $1$  (dash-dotted) and $1.5$  (dotted). We see  clearly that
the  impact  of  the  tadpole  correction is  to  broaden  the  bubble
wall.\label{fig:correctedbounce}}
\end{figure}

We note at this point that $G(u, u', m)$ is singular as $k \to 0$ (or,
equivalently,   $m    \to   2$).   Nonetheless,   the    integral   in
Eq.~\eqref{sol:corr:Green}  remains finite,  since  $G(u,  u', m)$  is
multiplied with  an odd function,  whereas the singularity  resides in
its even part. It is therefore useful to define
\begin{equation}
  G^{\rm odd}(u, u') \ \equiv \ \frac{1}{2}
  \Big(G(u, u', 2) - G(u, -\,u', 2) \Big) \; .
\end{equation}
Within  the  domain $0  \leq  u,  u' \leq  1$,  this  function can  be
expressed as
\begin{align}
  & G^{\rm odd}(u,u') \ = \ \vartheta(u - u') \, \frac{1}{32 \gamma} \,
  \frac{1 - u^2}{1 - u'^2} \bigg[ 2u' \big( 5 - 3 u'^2 \big)
  \nonumber \\
  & \qquad + 3 \big( 1 - u'^2 \big)^2 \ln \frac{1 + u'}{1 - u'} \bigg]
  + ( u \leftrightarrow u') \; .
\end{align}
Defining in addition
\begin{subequations}
  \begin{align}
    p_0(u) \ & = \ \gamma \! \int_{-1}^{1} \! \D u' \;
    \frac{u'}{1 - u'^2} \, G^{\rm odd}(u, u')
    \nonumber \\
    & = \ \frac{1 - u^2}{8} 
    \bigg[ \frac{2u}{1 - u^2} + \ln \frac{1 + u}{1 - u} \bigg] \; ,
    \\
    p_1(u) \ & = \ \gamma \! \int_{-1}^{1} \! \D u' \;
    u' G^{\rm odd}(u,u')
    \nonumber \\
    & = \ \frac{1 - u^2}{8} \ln \frac{1 + u}{1 - u} \; ,
    \\
    p_2(u) \ & = \ \gamma \! \int_{-1}^1 \! \D u' \;
    u'^3 G^{\rm odd}(u, u')
    \nonumber \\
    & = \ \frac{1 - u^2}{8}
    \bigg[ \ln \frac{1 + u}{1 - u} - \frac{4}{3} \, u \bigg] \; ,
  \end{align}
\end{subequations}
we find the result
\begin{align}
  \label{bounce:correction}
  & \delta \varphi(u) \ = \ - \, \frac{3 \lambda v}{16 \pi^2}
  \bigg[ 6 \bigg( \frac{8 \gamma^2}{m_{\chi}^2} \, N + 1 \bigg) p_0(u)
  \nonumber \\
  & \qquad + 5 \bigg( \frac{16 \gamma^2}{3 m_{\chi}^2} \, N +1 
  \bigg) p_1(u) - \bigg( \frac{2 \gamma^2}{m_{\chi}^2} \, N
  + \frac{\pi}{\sqrt{3}} \bigg) p_2(u) \bigg] \; .
\end{align}
In Fig.~\ref{fig:deltaphi}, we plot $\delta  \varphi$ as a function of
$\gamma(r-R)$ for a range of values of $N\gamma^2/m_{\chi}^2$.  We see
from Fig.~\ref{fig:correctedbounce}, which  plots the corrected bounce
$\varphi+\delta\varphi$ for  the same range,  that the impact  of this
correction is to lower the height  and broaden the width of the bubble
wall. We note that this behaviour is in qualitative agreement with the results of the self-consistent numerical analysis in Ref.~\cite{Bergner:2003au}, presented there for the pure $\lambda\Phi^4$ theory in $1+1$ dimensions.

\begin{figure}
  \centering
  \includegraphics[width=7.5cm]{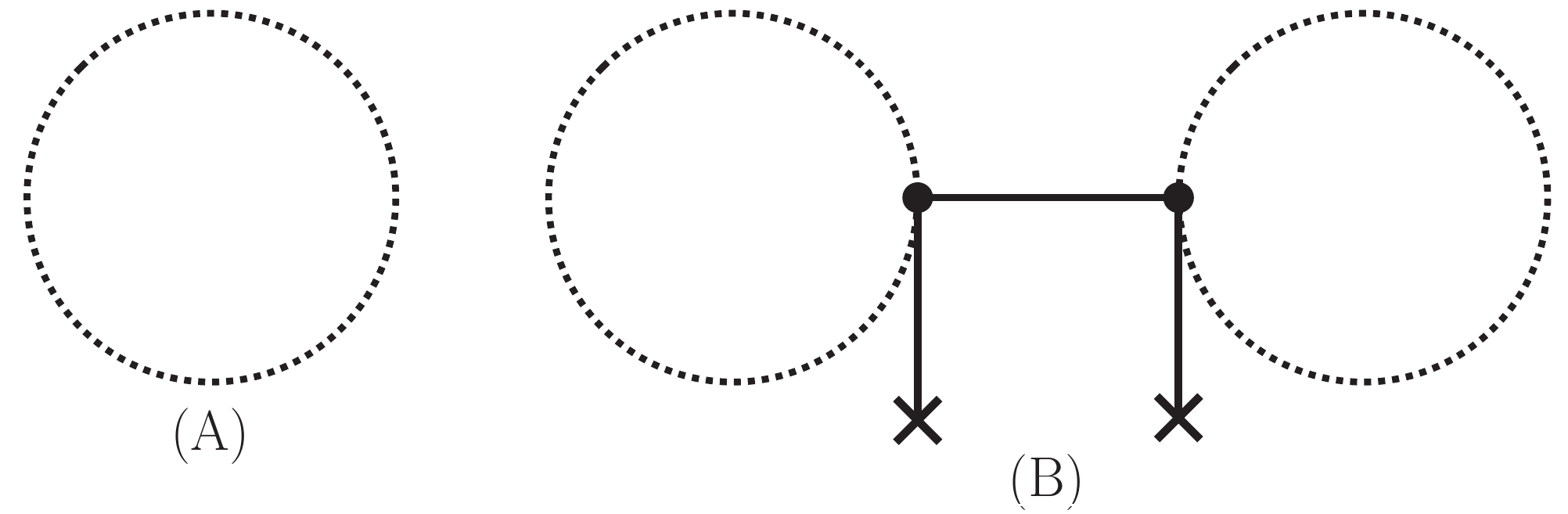}\\
  \vskip.4cm
  \includegraphics[width=4.8cm]{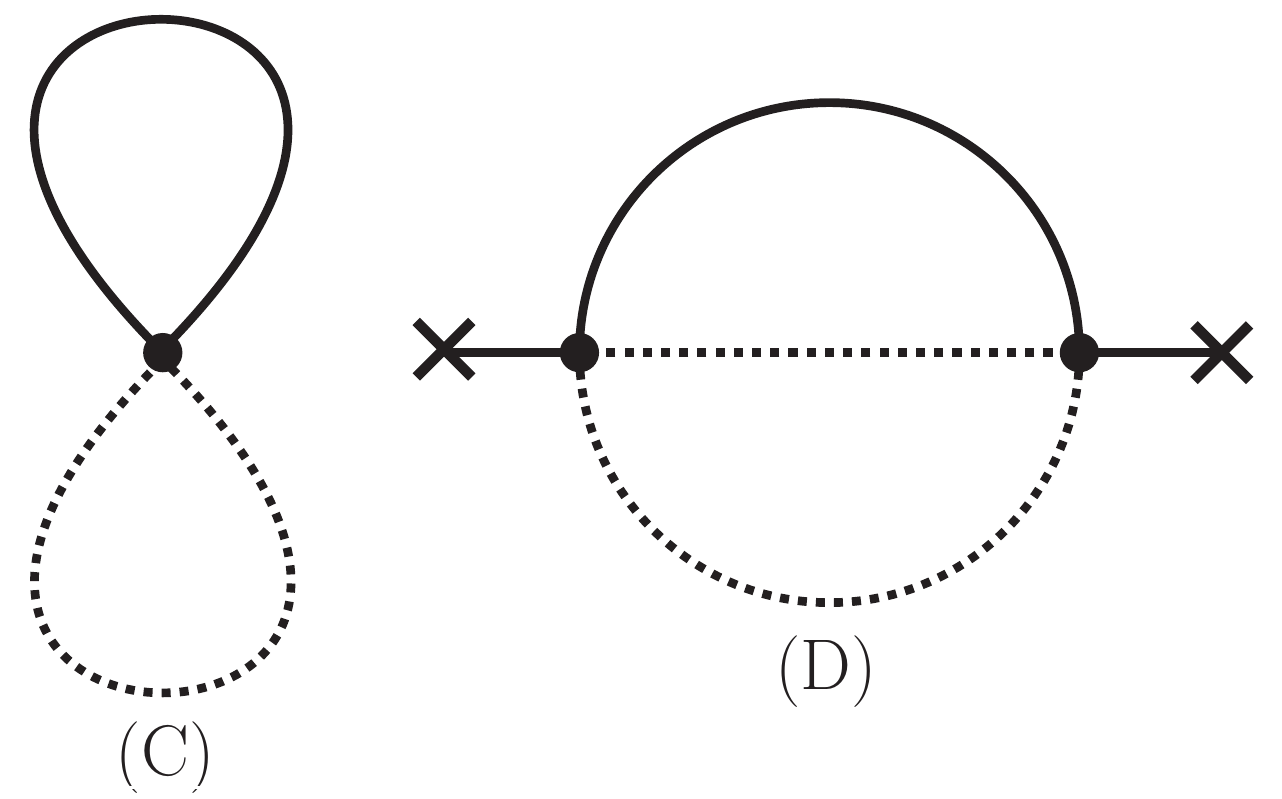}
  \caption{\label{fig:feyn:radcor}Diagrammatic    representation    of
various contributions  to the  effective action:  (A) is  the one-loop
term $B_\chi^{(1)}$, (B) is the ${\cal O}(\lambda^2 N^2)$ contribution to
$B^{(2)}$, and (C) and (D) are  ${\cal O}(\lambda^2 N)$ terms. Solid lines
represent   the  propagator   $G(\varphi;x,x^\prime)$,  dotted   lines
$S(\varphi;x,x^\prime)$.  Crosses  denote  insertions  of  the  bounce
$\varphi$.}
\end{figure}

Substituting Eq.~\eqref{bounce:correction}  into Eq.~\eqref{eq:B2}, we
find the correction to the bounce action
\begin{align}
  \label{delta:B:tree}
  B^{(2)}+B^{(2)}{}' \ & = \  \frac{1}{2} \! \int \! \D^4 x \; 
  \varphi(u) \, \Big( \Pi^R(u) + N \Sigma^R(u) \Big) \, \delta \varphi(u)
  \nonumber \\
  & = \ -\,\frac{B}{3} \bigg( \frac{3 \lambda}{16 \pi^2} \bigg)^{\! \! 2}
  \bigg[ \frac{291}{8} - \frac{37}{4} \, \frac{\pi}{\sqrt{3}}
  + \frac{5}{56} \, \frac{\pi^2}{3}
  \nonumber \\
  & \qquad \qquad + \bigg( \frac{667}{2}
  - \frac{2897}{42} \, \frac{\pi}{\sqrt{3}} \bigg)
  \frac{\gamma^2}{m_{\chi}^2} \, N
  \nonumber \\
  & \qquad \qquad + \frac{5829}{14} \,
  \frac{\gamma^4}{m_{\chi}^4} \, N^2 \bigg] \; .
\end{align}
In order  to obtain a  finite result for  Eq.~\eqref{delta:B:tree}, we
have added to $U_0$ the correction
\begin{equation}
  \delta U_0 \ = \ \frac{9}{4} \,
  \bigg( \frac{3 \lambda}{16 \pi^2} \bigg)^{\! \! 2} \, \gamma^2 v^2 \,
  \bigg(\frac{8 \gamma^2}{m_{\chi}^2} N + 1 \bigg)^{\! \! 2} \; ,
\end{equation}
ensuring that the potential continues to vanish in the false vacuum.

The  corrections  appearing   in  Eq.~\eqref{delta:B:tree}  should  be
compared  to the  renormalized logarithm  of the  determinants of  the
Klein-Gordon operators of the $\chi$ fields in the background given by $\varphi$,
which are given by
\begin{align}
  \label{chi:oneloop}
  B^{(1)}_\chi = - \, B \, \bigg( \frac{3 \lambda}{16 \pi^2} \bigg)
  \frac{2542}{15} \, \bigg[ \frac{\gamma^2}{m_{\chi}^2}
  + \mathcal{O} \big( \tfrac{\gamma^4}{m_{\chi}^4} \big) \bigg] N \; .
\end{align}
In  comparison,     the    leading     term     in
  Eq.~\eqref{delta:B:tree}  is suppressed  by a  factor $\sim  \lambda
  \mu^2 / m_\chi^2 / (16 \pi^2)$. The one-loop corrections $B^{(1)}$ and $B^{(1)}_{\chi}$ are both negative, thereby increasing the tunneling rate. It is interesting to note that, although the contribution $B^{(2)}$ to the tunneling action from the corrections to the bounce itself is positive, the net contribution of $B^{(2)}+B^{(2)}{}'$ is still negative, again increasing the tunneling rate.

In    Fig.~\ref{fig:feyn:radcor},    we   present    a    diagrammatic
representation of  the corrections  to the bounce  action. It  is also
 useful  in order  to see  that there  appear no  contributions of
${\cal  O}(\lambda^2  N^2)$  relative  to the  bounce  action  $B$  in
addition to those from $B^{(2)}$.  In order to avoid proliferation, we
only  show the  leading contributions  in $1/N$  for a  given type  of
diagram. At one-loop order, there is the vacuum bubble in terms of the
propagator  $S$ of  the  $\chi$ fields,  Fig.~\ref{fig:feyn:radcor}(A),
which gives  the contribution  ${\cal O}(\lambda  N)$ relative  to $B$
from  $B^{(1)}_\chi$  in Eq.~\eqref{chi:oneloop}.   On substituting
$\delta\varphi$  in the  form of  Eq.~\eqref{sol:corr:Green} into  the
action [Eq.~\eqref{eq:secondexp}], we see that the diagram corresponding
to  the ${\cal  O}(\lambda^2 N^2)$  term  in $B^{(2)}/B$  is given  by
Fig.~\ref{fig:feyn:radcor}(B),  where,  when  counting the  powers  of
$\lambda$, one should  note that each explicit  insertion of $\varphi$
contributes a factor of  $1/\sqrt\lambda$. Finally, at two-loop order,
there are the diagrams Figs.~\ref{fig:feyn:radcor}(C) and (D), which we
do  not compute  but yield  contributions of  ${\cal O}(\lambda^2  N)$
relative to  $B$.  These contributions  are therefore suppressed  by a
relative factor  of $1/N$ relative to the $\mathcal{O}(\lambda^2N^2)$ in $B^{(2)}/B$, as is  familiar from
the    standard    approximation scheme    known    as    the    $1/N$
expansion~\cite{'tHooft:1973jz}. We should remark that these arguments do not hold, of course, for the contribution to $B^{(2)}$ from the $\Phi$ tadpole, which we include here for completeness. The latter is formally the same order as other two-loop diagrams, involving only $\Phi$, that are not captured in the 1PI approximation employed here. This observation is true also of the Hartree approximation for the pure $\lambda\Phi^4$ theory analyzed numerically in Refs.~\cite{Bergner:2003au,Bergner:2003id,Baacke:2004xk,Baacke:2006kv}. Nevertheless, these additional two-loop diagrams remain subdominant compared to the $\mathcal{O}(\lambda^2 N)$ and $\mathcal{O}(\lambda^2N^2)$ contributions from the $\chi$ tadpole in Eq.~\eqref{delta:B:tree}. 

Finally, we  note that approximating
$\delta\varphi$  as   a  small  perturbation  to   $\varphi$  ,  using
Eq.~\eqref{sol:corr:Green}, requires for consistency that $6 N \lambda
\gamma^2/(m_\chi^2  \pi^2)  \ll 1$,  such  that  within the  range  of
validity    of    present    approximations,    we    cannot    obtain
$|B^{(2)}+B^{(2)}{}'|>|B^{(1)}|$.  Nevertheless  for large $N$, $B^{(2)}+B^{(2)}{}'$  can be the
dominant two-loop contribution to the effective action.


\section{Conclusions}
\label{sec:conc}

Within the  context of  $\lambda \Phi^4$ theory,  we have  described a
Green's function method for handling radiative effects on false vacuum
decay.  By  this  means  and   employing  the  thin-  and  planar-wall
approximations, we have  been able to calculate analytically  and in a
straightforward  manner   both  the  functional  determinant   of  the
quadratic fluctuations  about the  classical soliton  configuration and the first correction to the configuration itself.

This Green's  function method is  well suited to  numerical evaluation
and,  as a  consequence, should  be applicable  to potentials  of more
general form.  As such, we anticipate that it may be of particular use
when  the non-degeneracy  of minima  is purely  radiatively generated.
Examples of  the latter include  the spontaneous symmetry  breaking of
the  massless  CW   model~\cite{Coleman:1973jx}  or  the
instability  of  the  electroweak vacuum.   Other  applications  might
include the  calculation of corrections to  inflationary potentials in
the    time-dependent   inflaton    background,   for    instance   in
inflection-point   or    $A$-term   inflation~\cite{Allahverdi:2006iq,
Lyth:2006ec, Bueno  Sanchez:2006xk, Allahverdi:2006we},  which exploit
the  flat directions   and  saddle points   of  the   MSSM  potential.
Furthermore,  the  use  of  Green's  functions  naturally  admits  the
introduction of finite-temperature effects or extension to non-trivial
background spacetimes.

Green's   functions  have   proved  to   be  central   objects  within
perturbative calculations  throughout quantum  field theory and  it is
therefore unsurprising that  we find these suitable  to treat solitons
in $\lambda \Phi^4$  theory as well. We take this  as an encouragement
that   further   theoretically-   and   phenomenologically-interesting
systematic results on false vacuum decay may be within reach.


\begin{acknowledgments}

The authors would like to thank J\"{u}rgen Baacke, Daniel Litim and Holger Gies for helpful correspondence and discussions. The work  of P.M. is  supported by a University  Foundation Fellowship
(TUFF) from  the Technische  Universit\"{a}t M\"{u}nchen. The  work of
B.G. is  supported by the  Gottfried Wilhelm Leibniz Programme  of the
Deutsche  Forschungsgemeinschaft  (DFG).    Both  authors  acknowledge
support from the DFG cluster of excellence Origin and Structure of the
Universe.

\end{acknowledgments}


\appendix

\section{Zero-mode functional measure}
\label{app:zeromode}

In order to perform the functional integration over the zero modes, we
insert     four      copies     of     unity      in     Faddeev-Popov
form~\cite{Gervais:1974dc}:
\begin{equation}
  1 \ = \ \int \! \D y_{\mu} \; | \partial^{(y)}_{\mu} \! f(y_{\mu}) |
  \, \delta \big( f(y_{\mu}) \big) \; .
\end{equation}
Here, $\mu$ is not summed over and
\begin{equation}
  f(y_{\mu}) \ = \ \int \! \D^4 x \; \Phi(x - y)
  \partial^{(x)}_{\mu} \varphi(x - y) \ = \ B^{1/2} a_{\mu} \; ,
\end{equation}
where we recall that
\begin{equation}
  \Phi \ = \ \varphi + \sum_{i = 0}^4 a_i \phi_i + \phi_+ \; .
\end{equation}
It follows that
\begin{equation}
  \partial_{\mu}^{(y)} \! f(y_{\mu}) \ = \ - \! \int \! \D^4 x \;
  \big( \partial^{(x)}_{\mu} \varphi(x - y) \big)^2 \ = \ -\,B \; ,
\end{equation}
ignoring terms that are formally $\mathcal{O}(\hbar^{1/2})$. Thus,
\begin{equation}
  1 \ = \ B \! \int \! \D y_{\mu} \; \delta(B^{1/2} a_{\mu}) \
  = \ B^{1/2} \! \int \! \D y_{\mu} \; \delta(a_{\mu}) \; .
\end{equation}
We then have
\begin{align}
  \int \prod_{\mu = 1}^4 (2 \pi \hbar)^{-1/2} \D a_{\mu} \ &
  = \ \bigg( \frac{B}{2 \pi \hbar} \bigg)^{\! 2} \! \int \! \D^4 y
  \prod_{\mu = 1}^4 \int \mathrm{d} a_{\mu} \; \delta(a_{\mu})
  \nonumber \\
  & = \ VT \bigg( \frac{B}{2 \pi \hbar} \bigg)^{\! 2} \; .
\end{align}


\section{Green's function}
\label{app:greens}

In this appendix, we include the technical details of the calculations
outlined in Secs.~\ref{sec:greens} and~\ref{sec:corrb}. All functional
identities used in what follows may be found in Ref.~\cite{Abramowitz}.


\subsection{Expansion in hyperspherical harmonics}

In $d$ dimensions,  the Green's  function satisfies  the inhomogeneous
Klein-Gordon equation
\begin{equation}
  \big( - \Delta^{(d)} + U''(\varphi) \big) G^{(d)}(\varphi; x, x') \
  = \ \delta^{(d)}(x - x') \; ,
\end{equation}
where  $\delta^{(d)}(x  -  x')$  is   the  Dirac  delta  function  and
$\Delta^{(d)}$ is  the Laplacian. Given  the $O(d)$ invariance  of the
bounce  $\varphi$,   it  is  convenient  to   work  in  hyperspherical
coordinates, in which case the Laplacian takes the form
\begin{equation}
  \Delta^{(d)} \ = \ r^{1 - d} \partial_r r^{d - 1} \partial_r
  + \Delta_{S^{d - 1}} \; , 
\end{equation}
where $\Delta_{S^{d - 1}}$ is  the Laplace-Beltrami operator on the $d
- 1$ sphere.

We proceed by performing a partial-wave decomposition of the Green's function:
\begin{equation}
  G^{(d)}(\varphi; x, x')
  \\
  = \ \sum_{j \{\ell\}} G_j(\varphi; r, r')
  Y^*_{j\{\ell\}}(\mathbf{e}_{r'}) Y_{j\{\ell\}}(\mathbf{e}_{r}) \; ,
\end{equation}
where $x  = r \mathbf{e}_r$, $x'  = r' \mathbf{e}_{r'}$  and $Y_{j
  \{\ell\}}(\mathbf{e}_r)$  are  the   hyperspherical  harmonics  (see
e.g.~Ref.~\cite{Avery}), satisfying the eigenvalue equation
\begin{equation}
  \Delta_{S^{d-1}}Y_{j\{\ell\}}\ =\ -\,j(j+d-2)Y_{j\{\ell\}}\;,
\end{equation}
with $\{\ell\}  = \ell_1, \; \ell_2,  \; \dots, \; \ell_{d  - 2}$. The
hyperradial function $G_j(\varphi; r, r')$ satisfies
\begin{align}
  \label{eq:dradialdif}
  & \bigg[ -\, r^{1 - d} \, \frac{\D}{\D r} \, r^{d - 1} \,
  \frac{\D}{\D r} + \frac{j(j + d - 2)}{r^2}
  + U''(\varphi) \bigg] G_j(\varphi; r, r')
  \nonumber \\
  & \qquad = \ r'^{1 - d} \delta(r - r') \; .
\end{align}

Since, for each  $j$, the $\{\ell\}$ modes are degenerate,  we may use
the sum rule~\cite{Avery}
\begin{align}
  \label{eq:sumrule}
  & \sum_{\{\ell\}} Y^*_{j \{\ell\}}(\mathbf{e}_{r'})
  Y_{j \{\ell\}}(\mathbf{e}_{r})
  \\ \nonumber
  & \ = \ \frac{2}{(4 \pi)^{\kappa + \tfrac{1}{2}}} \,
  \frac{j + \kappa}{j + 2\kappa}
  \big( j + \kappa + \tfrac{1}{2} \big)_{\kappa + \tfrac{1}{2}}
  P_j^{\big(\kappa - \tfrac{1}{2}, \kappa - \tfrac{1}{2}\big)}(\cos \theta) \; ,
\end{align}
where $\kappa = d/2 -  1$, $\cos \theta = \mathbf{e}_r \cdot
\mathbf{e}_{r'}$,
\begin{equation}
  \label{eq:Pochhammer}
  (z)_n\ =\ \frac{\Gamma(z+n)}{\Gamma(z)}
\end{equation}
is  the Pochhammer  symbol and  the $P_j^{(\alpha,\beta)}(z)$  are the
Jacobi polynomials.

For $d  = 1$, $\kappa  = - \,  1/2$, $\cos \theta  \in \{-1,
+1\}$, and we have
\begin{gather}
  P_{j}^{(-1, -1)}(+1) \ = \ 0 \; ,
  \\
  P_{j}^{(-1, -1)}(-1) \ = \ \frac{\sin \pi j}{\pi j} \
  = \ \begin{cases} 1 \; , \quad j = 0
         \\
         0 \; , \quad j \neq 0
      \end{cases} \; .
\end{gather}
Hence, $G^{(1)}(\varphi;  x, x') =  G_0(\varphi; r, r')$, as  we would
expect.

For $d = 2$, $\kappa = 0$, and we have
\begin{equation}
  P_j^{\big(-\tfrac{1}{2}, -\tfrac{1}{2}\big)}(z) \ = \
  \frac{T_j(z)}{\sqrt{\pi} \big( j + \tfrac{1}{2} \big)_{\tfrac{1}{2}}} \; ,
\end{equation}
where $T_j(z)$ is the Chebyshev polynomial  of the first kind. We then
obtain
\begin{equation}
  G^{(2)}(\varphi; x, x') \ = \ \frac{1}{\pi} \sum_{j = 0}^{\infty}
  \cos j \theta \, G_{j}(\varphi; r, r') \; ,
\end{equation}
where we have  used the trigonometric form $T_j(\cos \theta)  = \cos j
\theta$.

For $d = 3$, $\kappa = 1/2$, and
\begin{equation}
  P_j^{(0, 0)}(z) \ = \ P_j(z) \; ,
\end{equation}
where  $P_j(z)$ are  the  Legendre polynomials.  Thus,  we obtain  the
familiar three-dimensional expansion
\begin{equation}
  G^{(3)}(\varphi; x, x') \ = \ \frac{1}{4\pi} \sum_{j = 0}^{\infty}
  (2 j + 1) P_j(\cos \theta) G_j(\varphi; r, r') \; .
\end{equation}

Finally, for $d = 4$, $\kappa = 1$, and
\begin{equation}
  P_j^{\big(\tfrac{1}{2}, \tfrac{1}{2}\big)}(z) \ = \ \frac{2}{\sqrt{\pi}} \, 
  \frac{j + 2}{\big( j + \tfrac{3}{2} \big)_{\tfrac{3}{2}} } \,
  U_j(z) \; ,
\end{equation}
where   $U_j(z)$  are   the  Chebyshev   polynomials  of   the  second
kind. Hence, we find
\begin{equation}
  G^{(4)}(\varphi; x, x') \ = \ \frac{1}{2 \pi^2}
  \sum_{j = 0}^{\infty} (j + 1) U_j(\cos \theta)
  G_j(\varphi; r, r') \; ,
\end{equation}
as appearing in Eq.~\eqref{eq:greenexp}.


\subsection{Continuum approximation}

In the coincident limit $x = x'$, $\cos \theta = 1$, and we have
\begin{gather}
  T_j(1) \ = \ 1 \; ,
  \quad
  P_j(1) \ = \ 1 \; ,
  \quad
  U_j(1) \ = \ j + 1 \; .
\end{gather}
Alternatively, in $d$ dimensions, we may use
\begin{equation}
  P_j^{(\alpha, \beta)}(1) \
  = \ \frac{(\alpha + 1)_j}{\Gamma(j + 1)} \; ,
\end{equation}
in Eq.~\eqref{eq:sumrule}, giving
\begin{align}
  & G^{(d)}(\varphi; x, x') \
  = \ \frac{2(4 \pi)^{-\tfrac{d - 1}{2}}}{\Gamma\big(\tfrac{d - 1}{2}\big)}
  \nonumber \\ 
  & \quad \times \sum_{j = 0}^{\infty}(j + d/2 - 1)\,
  \frac{\Gamma(j + d - 2)}{\Gamma(j + 1)}\, G_j(\varphi; r, r') \; .
\end{align}

Completing   the    square   in    the   centrifugal    potential   in
Eq.~\eqref{eq:dradialdif},  we make  the  following approximation  for
large $R$:
\begin{equation}
  \frac{j(j + d - 2)}{R^2} \ = \ \frac{(j + \kappa)^2}{R^2}
  - \frac{\lambda^2}{4 R^2} \ \approx \ \frac{(j + \kappa)^2}{R^2} \; ,
\end{equation}
where $\kappa=d/2-1$, as  before. We may then promote $(j  + \kappa) /
R$ to a continuous variable $k$, obtaining
\begin{subequations}
  \begin{align}
    G^{(2)}(\varphi; x, x) \ & = \ \frac{1}{\pi}
    \int_0^{\infty} \! \D k \; G(u, m) \; ,
    \\
    G^{(3)}(\varphi; x, x)\ & = \ \frac{1}{2 \pi}
    \int_0^{\infty} \! \D k \; k \, G(u, m) \; ,
    \\
    G^{(4)}(\varphi; x, x)\ & = \ \frac{1}{2 \pi^2}
    \int_0^{\infty} \! \D k \; k^2 \, G(u, m) \; ,
    \\
    \vdots \nonumber
    \\
    \label{eq:continuousind}
    G^{(d)}(\varphi; x, x) \ & = \ \frac{2(4 \pi)^{- \tfrac{d - 1}{2}}}
      {\Gamma \big( \tfrac{d - 1}{2} \big)} \! 
    \int_0^{\infty} \! \D k \; k^{d - 2} \, G(u, m) \; ,
  \end{align}
\end{subequations}
where   we    have   used    the   general   notation    employed   in
Sec.~\ref{sec:greens},  see  Eq.~\eqref{eq:Gum},  with  $m$  given  by
Eq.~\eqref{eq:continuum}. We  note that for  $d > 4$, we  have dropped
terms $\mathcal{O}(k / R)$ and higher within the integrand.


\subsection{Radial function}

For large  $R$, we neglect  the damping  term in  the  radial  equation
[Eq.~\eqref{eq:dradialdif}] and set  $r = R$  in the
centrifugal  potential  and  discontinuity, giving
\begin{equation}
  \label{eq:dradialsimple}
  \bigg[ - \, \frac{\D^2}{\D r^2} + \frac{j(j + d - 2)}{R^2}
  + U''(\varphi) \bigg] G_j(\varphi; r, r') \
  = \ \frac{\delta(r - r')}{R^{d - 1}} \; .
\end{equation}
Since   the   solution  depends   only   on   the  normalized   bounce
$u = \tanh[\gamma (r - R)]$, it is convenient to define
\begin{equation}
  G(u, u', m) \ \equiv \ R^{d - 1} \, G_j(\varphi; r, r') \; ,
\end{equation}
cf.~Sec.~\ref{sec:greens}. Equation~\eqref{eq:dradialsimple}  may then
be recast in the form
\begin{align}
  & \bigg[ \frac{\D}{\D u} \big( 1 - u^2) \frac{\D}{\D u}
  - \frac{m^2}{1 - u^2} + n(n + 1) \bigg] G(u, u', m)
  \nonumber \\
  & \qquad \qquad = \ - \, \gamma^{-1} \delta(u - u') \; ,
\end{align}
where
\begin{equation}
  n \ = \ 2 \; ,
  \qquad
  m \ = \ 2 \, \bigg( \! 1 + \frac{j(j + d - 2)}
    {4 \gamma^2R^2} \bigg)^{\!\!\tfrac{1}{2}} \; .
\end{equation}

Splitting around the discontinuity at $u = u'$, we decompose
\begin{align}
  G(u, u', m) \ & = \ \vartheta(u - u') G^>(u, u', m)
  \nonumber \\
  & \qquad \qquad 
  + \vartheta(u' - u) G^<(u, u', m) \; ,
\end{align}
where $G^{\gtrless}(u,  u', m)$ are  the solutions to  the homogeneous
equation
\begin{equation}
  \bigg[  \frac{\D}{\D u} \big(1 - u^2) \frac{\D}{\D u}
  - \frac{m^2}{1 - u^2} + n(n + 1) \bigg]
  G^{\gtrless}(u, u', m) \ = \ 0 \; .
\end{equation}
The latter  is the  associated Legendre  differential equation  and we
obtain the general solutions
\begin{equation}
  G^{\gtrless}(u, u', m) \ = \ A^{\gtrless} P_2^m(u)
  + B^{\gtrless} Q_2^m(u) \; ,
\end{equation}
where $P_n^m(z)$ and $Q_n^m(z)$  are the associated Legendre functions
of the first and second kind, respectively.

Matching around the  delta function in the  inhomogeneous equation, we
require
\begin{subequations}
  \begin{align}
    & \big( A^> - A^< \big) P_2^m(u') + \big( B^> - B^< \big) Q_2^m(u')
    \, = \, 0 \; ,
    \\
    & \big( A^> - A^< \big) \frac{\D }{\D u'} P_2^m(u')
    \nonumber \\
    & \quad + \big( B^> - B^< \big) \frac{\D }{\D u'} Q_2^m(u')
    = \, - \, \frac{1}{\gamma \big( 1 - u'^2 \big)} \; .
  \end{align}
\end{subequations}
Thus, we find
\begin{subequations}
  \begin{align}
     A^> - A^< \ = \ \frac{1}{\gamma \big( 1 - u'^2 \big)} \,
     \frac{Q_2^m(u')}{W[P_2^m(u'), Q_2^m(u')]} \; ,
     \\
     B^> - B^< \ = \ \frac{1}{\gamma \big( 1 - u'^2 \big)} \,
     \frac{P_2^m(u')}{W[Q_2^m(u'), P_2^m(u')]} \; ,
  \end{align}
\end{subequations}
where $W[P_n^m(z),Q_n^m(z)]$ is the Wronskian, having the explicit form
\begin{equation}
  W[P_n^m(z), Q_n^m(z)] \ = \ \frac{(n - m + 1)_{2m}}{1 - u'^2} \; ,
\end{equation}
with the  Pochhammer symbol defined in  Eq.~\eqref{eq:Pochhammer}.  We
also require the  boundary condition that $G(u, u', m)$  go to zero as
$u \to \pm \, 1$, giving
\begin{equation}
  \frac{A_>}{B_>} \ = \ - \, \frac{\pi}{2} \cot m \pi \; ,
  \qquad
  B_< \ = \ 0 \; .
\end{equation}

We may now solve for the remaining non-zero coefficients and obtain
\begin{equation}
  G^>(u, u', m) \ = \ \frac{\pi}{2\gamma} \,
  \frac{1}{\sin m \pi} \, P_2^{-m}(u) P_2^m(u') \; ,
\end{equation}
with $G^<(u, u', m) = G^>(u', u, m)$. Here, we have used the identity
\begin{equation}
  \frac{\pi (n - m + 1)_{2m}}{2\sin m \pi} \, \, P_n^{-m}(z) \
  = \ \frac{\pi}{2} \cot m \pi \, P_n^m(z) - Q_n^m(z) \; .
\end{equation}
Finally, we employ the representation
\begin{equation}
  P_n^m(z) \ = \ \bigg( \frac{z + 1}{z - 1} \bigg)^{\! \! \tfrac{m}{2}} \!
  (n - m + 1)_m \, P_n^{(-m, m)}(z)
\end{equation}
of the associated Legendre function of  the first kind in terms of the
Jacobi  polynomials. For  $n =  2$,  the polynomial  expansion of  the
latter terminates, and we have
\begin{align}
  \label{eq:polynomial}
  & P_2^{(\pm m, \mp m)}(z) \ = \ \frac{1}{2} \Big[(1 \pm m)(2 \pm m)
  \nonumber \\
  & \qquad \qquad - 3(2 \pm m)(1 - u) + 3(1 - u)^2 \Big] \; .
\end{align}
After  some algebraic  simplification,  we then  arrive  at the  final
analytic   solution,  as    presented   in   Eq.~\eqref{eq:Guupm}   of
Sec.~\ref{sec:greens}.


\subsection{Functional determinant}

The  normalized heat  kernel $\widetilde{\mathcal{K}}(\varphi;  z, z',
\mathbf{k} |  \tau)$, see~Sec.~\ref{sec:greens}, is given  in terms of
the inverse Laplace transform
\begin{equation}
  \widetilde{K}(\varphi; z, z', \mathbf{k} | \tau) \
  = \ \mathcal{L}_s^{-1}[\widetilde{G}(u, m)](\tau) \; ,
\end{equation}
where
\begin{equation}
  \widetilde{G}(u, m) \  = \ \frac{3}{2 \gamma m} \big( 1 - u^2 \big)
  \sum_{n = 1}^2 (-1)^n \frac{(n - 1 - u^2)}{m^2 - n^2} \; ,
\end{equation}
with
\begin{equation}
  m \ = \ 2 \, \bigg( 1 + \frac{k^2 + s}
    {4 \gamma^2} \bigg)^{\! \! \tfrac{1}{2}} \; .
\end{equation}

The inverse  Laplace transform  may be performed  by using  the shift,
scaling and division properties
\begin{subequations}
  \begin{gather}
    \mathcal{L}_{s}^{-1}[F(s + b)](\tau) \ = \ e^{- b\tau} f(\tau) \;,
    \\
    \mathcal{L}_{s}^{-1}[F(a s)](\tau) \ = \ \frac{1}{a} \, f(\tau / a)
    \; ,
    \\
    \mathcal{L}_{s}^{-1}[s^{-1} F(s)](\tau) \ = \
    \int_{0}^{\tau} \! \D \tau' \; f(\tau') \; ,
 \end{gather}
\end{subequations}
where  $f(\tau)  =  \mathcal{L}^{-1}_s[F(s)](\tau)$, as  well  as  the
elementary transformation
\begin{equation}
  \mathcal{L}_{s}^{-1}[s^{-z}](\tau) \, = \, \frac{\tau^{z - 1}}{\Gamma(z)}
  \; , \qquad \mathrm{Re} \, z \ > \ 0 \; .
\end{equation}
We find
\begin{equation}
  \mathcal{L}_s^{-1}[m^{-1}(m^2 - n^2)^{-1}](\tau)
  = \frac{\gamma^2}{n} \, 
  e^{[\gamma^2(n^2 - 4) - k^2]\tau}
  \mathrm{erf}(n\gamma\sqrt{\tau}) \; ,
\end{equation}
where
\begin{equation}
  \mathrm{erf}(z) \ = \ \frac{2}{\sqrt{\pi}} \!
  \int_0^z \! \D t \; e^{-t^2}
\end{equation}
is the error function. Hence, we have
\begin{align}
  & \widetilde{K}(\varphi; z, z', \mathbf{k} | \tau) \
  = \ - \, \frac{3}{2} \gamma \big( 1 - u^2 \big) e^{-k^2\tau}
  \nonumber \\
  & \times \sum_{n = 1}^2 (-1)^n \bigg( \frac{1 + u^2}{n} - 1 \bigg)
  e^{\gamma^2(n^2 - 4)\tau} \mathrm{erf}(n \gamma \sqrt{\tau}) \; .
\end{align}

Generalizing  to   $d$  dimensions,  using  the   continuum  limit  in
Eq.~\eqref{eq:continuousind},  the  correction  to the  bounce  action
arising from the functional determinant is therefore
\begin{align}
  & B^{(1)} \ = \ - \, \frac{3}{2}
 \frac{\Omega_d (4 \pi)^{- \tfrac{d - 1}{2}}}
 {\Gamma\big( \tfrac{d - 1}{2} \big)} \! 
 \int_{0}^{\infty} \! \D k \; k^{d - 2}
 \int_{0}^{\infty} \! \frac{\D \tau}{\tau} \, e^{-k^2\tau}
 \nonumber \\
 & \times \int_{0}^{\infty} \! \D r \; r^{d - 1} \, \gamma \big( 1 - u^2 \big)
 \nonumber \\
 & \times \sum_{n = 1}^2 (-1)^n \bigg( \frac{1 + u^2}{n} - 1 \bigg)
 e^{\gamma^2(n^2 - 4)\tau} \mathrm{erf}(n \gamma \sqrt{\tau}) \; ,
\end{align}
where $\Omega_d  = 2  \pi^{d/2} /  \Gamma(d / 2)$  is the  solid angle
subtended by the  $(d - 1)$ dimensional hypersphere.   The integral over
$r$ is dominated by $r \sim R$, such that (for $n = 1, 2$)
\begin{equation}
  (-1)^n \! \int_{0}^{\infty} \! \D r \; r^{d - 1} \,
  \gamma \big( 1 - u^2 \big) \bigg( \frac{1 + u^2}{n} - 1 \bigg) \
  \approx \ - \, \frac{2}{3} \, R^{d-1} \; .
\end{equation}
We are then left with
\begin{align}
  B^{(1)} \ & = \ \frac{\Omega_d (4 \pi)^{-\tfrac{d - 1}{2}}R^{d - 1}}
    {\Gamma\big(\tfrac{d - 1}{2}\big)} \!
  \int_0^{\infty} \! \D k \; k^{d - 2} \!
  \int_{0}^{\infty} \! \frac{\D\tau}{\tau} \; e^{-k^2\tau}
  \nonumber \\
  & \qquad \times \sum_{n = 1}^2 e^{\gamma^2(n^2 - 4)\tau}
  \mathrm{erf}(n \gamma \sqrt{\tau})\;,
\end{align}
cf.~the form presented in Ref.~\cite{Konoplich:1987yd}.

We  may now  proceed in  one of  two ways:  (i) performing  the $\tau$
integration first, we  must regularize the $k$  integral, for instance
by introducing  an ultraviolet cutoff $\Lambda$;  or (ii) performing
the  $k$  integral  first,  we  must  instead  regularize  the  $\tau$
integral.    The    latter    is    the    approach    presented    in
Ref.~\cite{Konoplich:1987yd},  which we  reproduce in  what follows  for
comparison.

\medskip

\noindent (i) Performing the $\tau$ integral first gives
\begin{align}
  B^{(1)} \ & = \  - \,
  \frac{2 \, \Omega_d (4 \pi)^{- \tfrac{d - 1}{2}} R^{d - 1}}
    {\Gamma \big( \tfrac{d - 1}{2} \big)}
  \int_{0}^{\Lambda} \! \D k \; k^{d - 2}
  \nonumber \\
  & \times \sum_{n = 1}^2 \mathrm{arcsinh} \,
  \frac{n \gamma}{\sqrt{k^2 - \gamma^2(n^2 - 4)}} \; .
\end{align}
Subsequently, performing the  $k$ integral for $d = 4$,  we obtain the
result in Eq.~\eqref{eq:B1final}.

\medskip

\noindent (ii) Instead, performing instead the $k$ integral first, we obtain
\begin{align}
  B^{(1)} \ & = \ \frac{1}{2} \, \Omega_d R^{d - 1}
  (4\pi)^{- \tfrac{d - 1}{2}} \! \int_{0}^{\infty} \! \! \D \tau \; 
  \tau^{- \tfrac{d + 1}{2}}
  \nonumber \\
  & \qquad \times \sum_{n = 1}^2 e^{\gamma^2(n^2 - 4)\tau}
  \mathrm{erf}(n \gamma \sqrt{\tau}) \; ,
\end{align}
which is regularized by introducing a large mass $M$ as follows:
\begin{align}
  B^{(1)} \ & = \ \frac{1}{2} \, \Omega_d R^{d - 1}
  (4\pi)^{- \tfrac{d - 1}{2}} \lim_{\epsilon \to 0} \frac{\D}{\D \epsilon}
  \frac{M^{2\epsilon}}{\Gamma(\epsilon)}
  \int_{0}^{\infty} \! \! \D \tau
  \nonumber \\
  & \qquad \times \tau^{- \tfrac{d + 1}{2} + \epsilon}
  \sum_{n = 1}^2e^{\gamma^2(n^2 - 4)\tau}
  \mathrm{erf}(n \gamma \sqrt{\tau}) \; .
\end{align}
We  may  proceed by  using  the  series  representation of  the  error
function
\begin{equation}
  \mathrm{erf}(z) \ = \ \frac{2}{\sqrt{\pi}} \, 
  e^{-z^2} \sum_{\ell = 0}^{\infty} \frac{2^{\ell}}{(2 \ell + 1)!!} \, 
  z^{2\ell + 1} \; ,
\end{equation}
where $!!$ denotes  the double factorial. The $\tau$  integral may now
be performed, and we obtain
\begin{align}
  & B^{(1)} \ = \ (\gamma R)^{d - 1} \Omega_d \pi^{- \tfrac{d}{2}} \,
  \lim_{\epsilon \to 0}\frac{\D}{\D \epsilon}
  \frac{1}{\Gamma(\epsilon)}
  \bigg( \frac{M^2}{4 \gamma^2} \bigg)^{\! \! \epsilon}
  \nonumber \\
  & \quad \times \sum_{n = 1}^2 \sum_{\ell = 0}^{\infty}
  \frac{2^{\ell} (n / 2)^{2 \ell + 1}}{(2 \ell + 1)!!} \,
  \Gamma(\epsilon + \ell + 1 - d / 2) \; .
\end{align}

Considering the derivative with respect to $\epsilon$, we have
\begin{align}
  & \frac{\D}{\D \epsilon} \,
  \frac{\Gamma(\epsilon + \ell + 1 - d / 2)}{\Gamma(\epsilon)}
  \bigg( \frac{M^2}{4 \gamma^2} \bigg)^{\! \! \epsilon}
  \nonumber \\
  & \quad
  = \ \frac{\Gamma( \epsilon + \ell + 1 - d / 2)}{\Gamma(\epsilon)}
  \bigg(\frac{M^2}{4 \gamma^2}\bigg)^{\! \! \epsilon}
  \nonumber \\
  & \quad \quad \times \bigg[ \ln \frac{M^2}{4 \gamma^2}
  - \psi(\epsilon) + \psi(\epsilon + \ell + 1 - d / 2) \bigg] \; ,
\end{align}
where $\psi(z)$ is  the digamma function. In order to  take the
limit $\epsilon  \to 0$ safely, we must  take note of the  poles occurring in
$\Gamma(z)$ and $\psi(z)$ for  non-positive integers. Such poles occur
in even dimensions for $\ell = 0,\, 1,\, \dots,\, d - 3$.

After  treating the  limit  $\epsilon  \to 0$,  we  find  for $d$  odd
(including $d = 1$)
\begin{align}
  B^{(1)} \ & = \ - \, (\gamma R)^{d - 1} \Omega_d \pi^{- \tfrac{d}{2}}
  \nonumber \\
  & \qquad \times \sum_{n = 1}^2 \sum_{\ell = 0}^{\infty}
  \frac{2^{\ell}(n / 2)^{2 \ell + 1}}{(2 \ell + 1)!!} \,
  \Gamma(\ell + 1 - d / 2) \; .
\end{align}
On the other hand, for $d$ even, we find
\begin{align}
  & B^{(1)} \ = \ - \, (\gamma R)^{d - 1} \Omega_d \pi^{- \tfrac{d}{2}}
  \nonumber \\
  & \times \sum_{n = 1}^2 \Bigg[ \sum_{\ell = d - 2}^{\infty}
  \frac{2^{\ell}(n / 2)^{2 \ell + 1}}{(2 \ell + 1)!!} \,
  \Gamma(\ell + 1 - d / 2)
  \nonumber \\
  & + \sum_{\ell = 0}^{d - 3} \frac{2^{\ell}(n / 2)^{2 \ell + 1}}
  {(2 \ell + 1)!!} \,
  \frac{(-1)^{d / 2 - \ell - 1}}{(d / 2 - \ell - 1)!}
  \bigg(\ln \frac{M^2}{4 \gamma^2} + H_{d / 2 - \ell - 1} \bigg) \Bigg] \; ,
\end{align}
where
\begin{equation}
  H_n \ = \ \sum_{k = 1}^n \frac{1}{k}
\end{equation}
are the harmonic numbers, which  we have supplemented with $H_0 \equiv
0$ for notational simplicity.

For $d=4$, we then obtain
\begin{align}
  & B^{(1)} \ = \ - 2 R^3 \gamma^3 \sum_{n = 1}^2
  \Bigg[\sum_{\ell = 2}^{\infty}
  \frac{2^{\ell}(n / 2)^{2 \ell + 1}}{(2 \ell + 1)!!}(\ell - 2)!
  \nonumber \\
  & + \sum_{\ell = 0}^1 \frac{2^{\ell}(n / 2)^{2 \ell + 1}}{(2 \ell + 1)!!} \,
  \frac{(-1)^{1 - \ell}}{(1 - \ell)!}
  \bigg(\ln \frac{M^2}{4 \gamma^2} + H_{1 - \ell} \bigg) \Bigg] \; .
\end{align}
Lastly, performing the summations, we arrive at the result
\begin{equation}
  \label{eq:BK}
  B^{(1)} \ = \ - \, B \, \bigg( \frac{3 \lambda}{16 \pi^2} \bigg)
  \bigg( \frac{\pi}{3 \sqrt{3}} - 2 + \ln \frac{4 \gamma^2}{M^2} \bigg)
  \; .
\end{equation}

Defining   the  counterterms   in   the  proper-time   representation,
see Ref.~\cite{Konoplich:1987yd},   and  fixing   the  renormalization
conditions as in Eq.~\eqref{eq:renormconds}, we find the counterterms
\begin{subequations}
  \begin{align}
    \delta m^2 \ & = \ \frac{\lambda \gamma^2}{16 \pi^2}
    \bigg( \ln \frac{4 \gamma^2}{M^2} + 29 \bigg) \; ,
    \\
    \delta \lambda \ & = \ - \, \frac{3 \lambda^2}{32 \pi^2}
    \bigg( \ln \frac{4 \gamma^2}{M^2} + 3 \bigg) \; ,
  \end{align}
\end{subequations}
giving
\begin{equation}
  \delta B^{(1)} \ = \ B \, \bigg( \frac{3 \lambda}{16 \pi^2} \bigg)
  \, \bigg( \ln \frac{4 \gamma^2}{M^2} - 23 \bigg) \; .
\end{equation}
Adding these to  Eq.~\eqref{eq:BK}, we obtain agreement with
Eq.~\eqref{eq:B1final}.


\section{Renormalization of the $\bm{N}$-field model}
\label{app:Scalc}

In this final appendix, we highlight the main technical details of the
derivation of the Green's function  and corrections to the bounce from
the $\chi$ fields.

Proceeding    as    for    the   isolated    $\varphi$    case,    see
Sec.~\ref{sec:greens},   the  renormalization   is  fixed   using  the
CW effective  potential~\cite{Coleman:1973jx}, evaluated
in  a homogeneous  false vacuum.   The renormalization  conditions are
then
\begin{subequations}
  \begin{gather}
    \frac{\partial^2 U_{\mathrm{eff}}}{\partial \varphi^2}
      \bigg|_{\varphi = v , \, \chi_i = 0} \ = \ 4 \gamma^2 \;,
    \\
    \frac{\partial^2 U_{\mathrm{eff}}}{\partial \chi_i^2}
      \bigg|_{\varphi = v , \, \chi_i = 0} \ = \ 6 \gamma^2 + m_{\chi}^2 \; ,
    \\
    \frac{\partial^4 U_{\mathrm{eff}}}{\partial \varphi^4}
      \bigg|_{\varphi = v , \, \chi_i = 0} \ = \ \lambda \; ,
    \\
    \frac{\partial^4 U_{\mathrm{eff}}}{\partial \varphi^2 \partial \chi_i^2}
    \bigg|_{\varphi = v , \, \chi_i = 0} \ = \ \lambda \; ,
    \\
    \frac{\partial^4 U_{\mathrm{eff}}}{\partial \chi_i^2 \partial \chi_j^2}
    \bigg|_{\varphi = v , \, \chi_i = 0} \ = \ 0 \; ,     
  \end{gather}
\end{subequations}
where the effective potential is
\begin{align}
  \label{eq:CWU}
  & U_{\mathrm{eff}} \ = \ U(\varphi, \chi) + \delta U(\varphi, \chi)
  \nonumber \\ 
  & \ + \frac{N-1}{4 \pi^2} \int_0^{\Lambda} \! \D k \; k^2
  \bigg( \sqrt{k^2 + M_{\chi}^2} - k \bigg)
  \nonumber \\
  & \ + \frac{1}{4 \pi^2} \int_0^{\Lambda} \! \D k \; k^2
  \bigg(\sqrt{k^2+M_+^2}+\sqrt{k^2+M_-^2} - 2k \bigg)\;,
\end{align}
with
\begin{subequations}
\begin{align}
M_{\chi}^2\ &=\ m_{\chi}^2+\frac{\lambda}{2}\,\varphi^2\;,\\
M_{\varphi}^2\ &=\ -\,2\gamma^2+\frac{\lambda}{2}\,\varphi^2+\frac{\lambda}{2}\,\chi_i^2\;,\\
M_{\pm}^2\ &= \ \frac{M_{\varphi}^2+M_{\chi}^2}{2}\pm\bigg[\bigg(\frac{M_{\varphi}^2-M_{\chi}^2}{2}\bigg)^{\!2}+\lambda^2\varphi^2\chi_i^2\bigg]^{\!1/2}\;,
\end{align}
\end{subequations}
and
\begin{subequations}
  \label{eq:CWUparts}
  \begin{align}
    &U(\varphi, \chi) \  = \ - \frac{1}{2!} \, \mu^2 \varphi^2
    + \frac{1}{2!} \, m_{\chi}^2 \chi_i^2
    \nonumber \\
    &\qquad + \frac{1}{4!} \, \lambda \varphi^4
    + \frac{1}{4} \, \lambda \varphi^2 \chi_i^2 \; ,
    \\
    &\delta U(\varphi, \chi) \  = \
    + \frac{1}{2!} \, \delta m_{\varphi}^2 \varphi^2
    + \frac{1}{2!}\,\delta m_{\chi}^2 \chi_i^2
    \nonumber \\
    &\qquad 
    + \frac{1}{4!} \, \delta \lambda_{\varphi} \varphi^4
    + \frac{1}{4} \, \delta \lambda_{\varphi\chi} \varphi^2 \chi_i^2
	+ \frac{1}{4}\,\delta \lambda_{\chi} \chi_i^2\chi_j^2    
     \; .
  \end{align}
\end{subequations}
In Eqs.~\eqref{eq:CWU}--\eqref{eq:CWUparts},  the summations over $i,j
= 1, \dots, N$ have been left implicit for notational convenience.

Solving the resulting system, we obtain the set of counterterms
\begin{subequations}
  \begin{align}
  & \delta m_{\chi}^2 \ = \ - \, \frac{\lambda \gamma^2}{16 \pi^2}
  \bigg(\frac{\Lambda^2}{\gamma^2} - \ln \frac{\gamma^2}{\Lambda^2}
  - 13\nonumber\\& \quad  +\frac{216\gamma^2}{m_{\chi}^2+6\gamma^2}-\frac{360\gamma^2}{m_{\chi}^2+2\gamma^2}\ln\frac{6\gamma^2+m_{\chi}^2}{4\gamma^2}\bigg) \; ,
  \\
  & \delta m_{\varphi}^2 \ = \ - \, \frac{\lambda \gamma^2}{16 \pi^2}
  \bigg[\big( N + 1 \big) \bigg( \frac{\Lambda^2}{\gamma^2} -30 \bigg)
  - \bigg( \ln \frac{\gamma^2}{\Lambda^2} + 1 \bigg)
  \nonumber \\
  & \ + N \frac{m_{\chi}^2}{2 \gamma^2}
  \bigg( \ln \frac{6 \gamma^2 + m_{\chi}^2}{4 \Lambda^2} + 1 \bigg)
  + 27 N \bigg( \frac{m_{\chi}^2 + 2 \gamma^2}
    {m^2_{\chi} + 6\gamma^2} \bigg)^{\! \! 2} \bigg] \; ,
  \\
  & \delta \lambda_{\varphi} \ = \ - \, \frac{3 \lambda^2}{32 \pi^2}
  \bigg[ \ln \frac{\gamma^2}{\Lambda^2} + 5(N + 1)
  \nonumber \\
  & \ + N \ln \frac{6\gamma^2 + m_{\chi}^2}{4 \Lambda^2}
  - 3 N \bigg( \frac{m_{\chi}^2 + 2\gamma^2}
    {m^2_{\chi} + 6\gamma^2} \bigg)^{\! \! 2} \bigg] \; ,\\
    & \delta \lambda_{\varphi\chi} \ = \ - \, \frac{\lambda^2}{32 \pi^2}
  \bigg( \ln \frac{\gamma^2}{\Lambda^2}+4\ln\frac{6\gamma^2+m_{\chi}^2}{4\Lambda^2}\nonumber\\& \quad +\frac{136\gamma^2}{m_{\chi}^2+2\gamma^2}\ln\frac{6\gamma^2+m_{\chi}^2}{4\gamma^2} +9\,\frac{m_{\chi}^2-2\gamma^2}{m_{\chi}^2+6\gamma^2} \bigg) \; ,
  \displaybreak\\
  &\delta\lambda_{\chi}\ =\ -\,\frac{\lambda^2}{32\pi^2}\bigg(\frac{1}{2}\ln\frac{\gamma^2}{\Lambda^2}+\frac{\big(m_{\chi}^2-10\gamma^2\big)^2+432\gamma^4}{\big(m_{\chi}^2+2\gamma^2\big)^{2}}\nonumber\\&\quad +24\gamma^2\frac{\big(m_{\chi}^2-2\gamma^2\big)^2-112\gamma^4}{\big(m_{\chi}^2+2\gamma^2\big)^3}\ln\frac{6\gamma^2+m_{\chi}^2}{4\gamma^2}\bigg)\;.
  \end{align}
\end{subequations}

Proceeding  as  for  $\varphi$,  we find  the  unrenormalized  tadpole
contribution of the $\chi$ fields
\begin{align}
  & \Sigma(u) \ = \ \frac{\gamma^2\lambda}{16 \pi^2}
  \Bigg[ \frac{\Lambda^2}{\gamma^2}
  + \frac{6 \gamma^2 + m_{\chi}^2}{2 \gamma^2}
  \bigg( \ln \frac{6 \gamma^2 + m_{\chi}^2}{4 \Lambda^2} + 1 \bigg)
  \nonumber \\
  & -3\big(1-u^2\big)\ln \frac{6 \gamma^2 + m_{\chi}^2}{4 \Lambda^2}\nonumber\\
   & - 6 \big( 1 - u^2 \big) \sum_{n = 1}^2 (-1)^n \big( n - 1 - u^2 \big)
  \nonumber \\
  & \times \bigg( \frac{6 \gamma^2 + m_{\chi}^2}{n^2 \gamma^2}
  - 1\bigg)^{\! \! \tfrac{1}{2}}
  \mathrm{arccot} \bigg( \frac{6 \gamma^2 + m_{\chi}^2}
    {n^2 \gamma^2} - 1 \bigg)^{\! \! \tfrac{1}{2}} \Bigg] \; .
\end{align}
After adding the counterterms, we obtain
\begin{align}
  \label{GFren:chi}
  & \Sigma^R(u) \ = \ \frac{3 \gamma^2\lambda }{16 \pi^2}
  \Bigg[ 11 - 5 u^2 - 3 \big( 3 - u^2 \big)
  \bigg( \frac{m_{\chi}^2 + 2\gamma^2}{m_{\chi}^2 + 6\gamma^2}\bigg)^{\! \! 2}
  \nonumber \\
  & - 2 \big( 1 - u^2 \big) \sum_{n = 1}^2 (-1)^n \big( n - 1 - u^2 \big)
  \nonumber \\
  & \times \bigg(\frac{6 \gamma^2 + m_{\chi}^2}{n^2 \gamma^2}
  - 1 \bigg)^{\! \! \tfrac{1}{2}}
  \mathrm{arccot} \bigg(
  \frac{6 \gamma^2 + m_{\chi}^2}{n^2 \gamma^2} - 1 \bigg)^{\! \! \tfrac{1}{2}}
  \Bigg] \; .
\end{align}
We note that  the expression in Eq.~\eqref{GFren:chi}  agrees with the
renormalized tadpole contribution from $\Phi$ in Eq.~\eqref{eq:phitad}
for $m_\chi^2 = - \mu^2$, as we would expect. Assuming $m_{\chi}^2 \gg
\gamma^2$,  we may  expand Eq.~\eqref{GFren:chi}  to leading  order in
$\gamma^2 / m_{\chi}^2$, giving Eq.~\eqref{eq:SR}.

The one-loop correction to the bounce action from the determinant over
the quadratic fluctuations in the $\chi$ fields is given by
\begin{align}
  B^{(1)}_{\chi} \ & = \ - \, \frac{N}{2} \int_0^{\Lambda} \! \D k \; k^2
  \int_0^{\infty} \! \frac{\D \tau}{\tau} \int_0^{\infty} \! \D r \; r^3
  \nonumber \\
  & \qquad \times \mathcal{L}^{-1}_s[\widetilde{S}(u, m)](\tau) \; ,
\end{align}
where    $\widetilde{S}(u,    m)$   is  obtained from    Eqs.~\eqref{eq:Gum}
and~\eqref{eq:Gtilde} with
\begin{align}
  m \ = \ \sqrt{6} \bigg( 1 + \frac{k^2 + s + m_{\chi}^2}
    {6 \gamma^2} \bigg)^{\! \! \tfrac{1}{2}} \; .
\end{align}
Continuing as in Sec.~\ref{sec:greens}, we find
\begin{align}
  B^{(1)}_{\chi} \ = \ & - N \frac{R^3 \gamma^3}{2}
  \bigg[ 3 \, \frac{\Lambda^2}{\gamma^2}
  + 3 \, \frac{m_{\chi}^2 + 4\gamma^2}{2 \gamma^2} \, 
  \ln \frac{6 \gamma^2 + m_{\chi}^2}{4 \Lambda^2}
  \nonumber \\
  & - \frac{m_{\chi}^2 + 2 \gamma^2}{2 \gamma^2}
  + \frac{2}{3} \sum_{n = 1}^2 n^3
  \bigg( \frac{6 \gamma^2 + m_{\chi}^2}{n^2 \gamma^2}
  - 1\bigg)^{\! \! \tfrac{3}{2}}
 \nonumber \\
 & \times \mathrm{arccot} \bigg(
 \frac{6 \gamma^2 + m_{\chi}^2}{n^2 \gamma^2}
 - 1\bigg)^{\! \! \tfrac{1}{2}} \bigg] \; .
\end{align}
Adding the counterterm
\begin{align}
  \delta B^{(1)}_{\chi} \ = \ & \frac{3}{2} \, N R^3 \gamma^3
  \bigg[ \frac{\Lambda^2}{\gamma^2} - 20 + \frac{m_{\chi}^2}{2 \gamma^2}
  + 21 \bigg( \frac{m_{\chi}^2 + 2 \gamma^2}{m_{\chi}^2 + 6\gamma^2}
  \bigg)^{\! \! 2}
  \nonumber \\
  & + \frac{m_{\chi}^2 + 4 \gamma^2}{2 \gamma^2}
  \ln \frac{6 \gamma^2 + m_{\chi}^2}{4 \Lambda^2} \bigg] \; ,
\end{align}
obtained in analogy with Eq.~\eqref{eq:deltaB}, we find
\begin{align}
  \label{eq:B1chifull}
  & B^{(1)}_{\chi} \ = \ - \, N \, \frac{R^3 \gamma^3}{2}
  \bigg[ 63 - 4 \, \frac{m_{\chi}^2 + 2 \gamma^2}{2 \gamma^2}
  - 63 \bigg( \frac{m_{\chi}^2 + 2 \gamma^2}{m_{\chi}^2 + 6\gamma^2}
  \bigg)^{\! \! 2}
  \nonumber \\
  & + \frac{2}{3} \sum_{n = 1}^2 n^3
  \bigg(\frac{6 \gamma^2 + m_{\chi}^2}{n^2 \gamma^2} - 1
  \bigg)^{\! \! \tfrac{3}{2}}
  \mathrm{arccot} \bigg(
  \frac{6 \gamma^2 + m_{\chi}^2}{n^2 \gamma^2}
  - 1\bigg)^{\! \! \tfrac{1}{2}} \bigg] \; .
\end{align}
The  result  in  Eq.~\eqref{eq:B1chifull}  reduces to  that  found  in
Eq.~\eqref{eq:B1final} for $m_{\chi}^2 = - 2 \gamma^2$ and $N = 1$, as
we would expect. Instead, taking  $m_{\chi}^2 \gg \gamma^2$, we obtain
the expression in Eq.~\eqref{chi:oneloop}.


\end{document}